\documentclass[twocolumn,prb,showpacs]{revtex4}
\usepackage{graphicx}
\usepackage{dcolumn}
\usepackage{bm}
\usepackage{amsmath} 
\usepackage{color}
\def\simge{\lower0.7ex\hbox{$\ \overset{>}{\sim}\ $}}
\def\simle{\lower0.7ex\hbox{$\ \overset{<}{\sim}\ $}}
\def\doteqa{\mathrel{\offinterlineskip\setbox0\hbox{$=$}%
\vbox to\ht0{\vss\hsize\wd0%
\hbox to\wd0{\hfil$\smash.$\hfil\hfil\hfil\hfil\hfil}%
\vskip0.1zw\copy0\vskip0.1zw%
\hbox to\wd0{\hfil\hfil\hfil\hfil\hfil$\smash.$\hfil}%
\vss}}}

\begin{document}

\preprint{APS/123-QED}
\title{Strong coupling superconductivity mediated by three-dimensional
anharmonic phonons}

\author{K. Hattori}
\email{hattori@issp.u-tokyo.ac.jp}
\author{H. Tsunetsugu}%
\affiliation{%
Institute for Solid State Physics, University of Tokyo, Kashiwanoha 5-1-5, Kashiwa, Chiba 277-8581, Japan
}%

\date{\today}

\begin{abstract}
We investigate 
three-dimensional anharmonic phonons in tetrahedral symmetry and 
superconductivity mediated by these phonons. 
Three-dimensional anharmonic phonon
spectra are calculated directly by solving Schr\"odinger equation and 
the superconducting transition temperature is determined by using the
theory of strong coupling superconductivity assuming an isotropic gap
function. With increasing the third-order anharmonicity $b$ of the
 tetrahedral ion potential, we find a crossover in the energy spectrum 
to a quantum tunneling regime. 
We obtain strongly enhanced transition temperatures around 
the crossover point. We also investigate the
 anharmonic effects on the Debye-Waller factor, the
 phonon spectral functions and the density profile, 
as a function of the anharmonicity $b$ and temperature. The isomorphic 
first-order transition observed in KOs$_2$O$_6$ is discussed in terms of 
the first excited state energy $\Delta$,
and the coupling constant $\lambda$ in the strong coupling theory of
superconductivity. 
Our results suggest the decrease in $\lambda$ and
increase in $\Delta$ below the first order transition temperature. 
We point out that the change in the oscillation amplitude $\langle
 x^2\rangle$ and $\langle xyz \rangle$ characterizes this isomorphic 
transition. The chemical trends of the
superconducting transition temperature, $\lambda$, and $\Delta$ in the
$\beta$-pyrochlore compounds are also discussed.
\end{abstract}

\maketitle

\section{Introduction}\label{intro}
Recently, various low-energy properties arising from anharmonic ion 
oscillations 
 have attracted much attention. An ion located at the center of an over-sized
cage oscillates with large amplitude and thus the anharmonic terms 
 in the potential energy play an important role. Indeed, these anharmonic oscillations are observed in $\beta$-pyrochlore,\cite{HiroiUnprecedented,Hiroi2ndAno,SummaryHiroi,Hasegawa}
filled-skutterudite,\cite{Goto,Kaneko,Kotegawa} and clathrate compounds.
\cite{Sales, Zerec, Matsumoto} In metallic systems, such anharmonic 
oscillations interact with conduction electrons, and  due to their 
large amplitude, the electron-phonon coupling constant becomes large. 
 The large anharmonicity and the strong electron-phonon coupling lead unusual relaxation
of conduction electrons,\cite{SummaryHiroi,ThermalCondKasahara}
anomalous nuclear magnetic relaxation time,\cite{Yoshida,Nakai} the sound velocity anomalies,\cite{Goto,Nemoto,Ishii,Nakanishi} and 
the strong coupling superconductivity.\cite{BattloggSC,IsotropicGappSCSC,peneShimono, SummaryHiroi,Nagao,Thalmeier,Chang}

These anharmonic oscillations in these
systems have also been studied theoretically. 
Dahm and Ueda discussed the anomalous temperature dependence of the
resistivity\cite{HiroiUnprecedented, Hiroi2ndAno,SummaryHiroi}
and the NMR relaxation time\cite{Yoshida} observed in KOs$_2$O$_6$ by
using a single-site anharmonic phonon model, employing the self-consistent Gaussian
approximation for the quartic term of the ion 
displacement.\cite{DahmUeda} Recently, Yamakage and Kuramoto extended
this approach to the lattice problem in the same level of Gaussian 
approximation.\cite{Yamakage} These works can
explain the softening of optical-phonon frequency as temperature
decreases observed in many compounds.\cite{Hasegawa,
Matsumoto,NeutronSasai, Galati, Mukta}

The anharmonic ion oscillations had been discussed also as a possible 
mechanism of high transition temperature superconductivity, particularly for
high $T_{\rm c}$ cuprates.\cite{Plakida, Hardy, Crespi,MahanSofo}
As for the systems of anharmonic ion oscillations in over-sized cages,
importance of contributions of low-energy   
Pr ion oscillations to the superconductivity is pointed
out in PrOs$_4$Sb$_{12}$.\cite{Thalmeier} 
Recently, Chang, {\it et al}.,\cite{Chang} discussed the superconductivity in
KOs$_2$O$_6$ 
using strong coupling theory of $s$-wave superconductivity\cite{Manalo}
 also employing the Gaussian
 approximation\cite{DahmUeda} for anharmonicity. 

In this paper, we focus on the $\beta$-pyrochlore compounds $A$Os$_2$O$_6$
($A$=K, Rb, or Cs). Monovalent $A$-cations are located at the center of the 
Os$_{12}$O$_{18}$ cages and form a diamond lattice structure. 
Lattice dynamics was investigated by neutron-scattering experiments,\cite{NeutronSasai,Galati,Mukta} and 
a low-energy optical phonon is observed at about 3 meV in the
K compound. This is basically K-cation oscillations, and 
the same phonon is also observed at around 5-7 meV in the 
Rb and Cs compounds. The root-mean-square amplitude of 
the K oscillation turns out to be 0.12-0.14 \AA\  
at zero temperature from the elastic neutron
scattering\cite{NeutronSasai,Galati} while much smaller values are 
reported for Rb and Cs compounds.\cite{NeutronSasai}

The first-principle
calculations indicate that the K ion potential
 has large anharmonicity and is very 
shallow along [111] and three other equivalent directions.\cite{BandCal}  
It is also calculated that the first excited state has the excitation 
energy of $\sim$8 K and K-cation oscillation amplitude is as large as 1 \AA\ 
 at zero temperature. These values are quite different from the
 experimental data, indicating that some parameters in the ion potential 
are not so realistic. In the present paper, we will first systematically 
analyze the effects of anharmonicity of the ion potential on the ion
dynamics in KOs$_2$O$_6$.

The $\beta$-pyrochlore compounds reveal superconducting phase transition 
and the transition temperature $T_{\rm c}$'s 
are 9.6 K, 6.3 K and 3.3 K for $A$=K, Rb, and Cs, respectively. These values of
$T_{\rm c }$ are inversely related to the $A$-cation size, {\it i.e.,} $T_{\rm c}$
is the highest for the smallest ion: potassium. The symmetry of the gap
function is considered to be fully gapped
$s$-wave.\cite{Yoshida,IsotropicGappSCSC,peneShimono}
It has been also shown that the electron-phonon
coupling is large.\cite{SummaryHiroi,IsotropicGappSCSC,Nagao}
Thus, it is expected that the conduction electrons on the cages strongly 
interact with the anharmonic oscillations of ion inside the cage in these 
systems.

 For KOs$_2$O$_6$, in addition to superconducting transition,  there exists
 a first-order structural transition at $T_p=7.5$ K.  
At this transition, no sign of symmetry breaking has been 
observed.\cite{SummaryHiroi,Hasegawa} 
The oscillation of K-cation seems less anharmonic 
below $T_p$ as indicated by electrical resistivity,
 specific heat jump at $T_{\rm c}$ in magnetic fields and
the mean free path estimated from the upper critical field 
$H_{c2}$.\cite{SummaryHiroi}
Recently, we have proposed 
that this is driven by a sudden change in K-cation oscillation 
amplitude driven by intersite ion interactions.\cite{Hat} 
The amplitude of the low-energy excited state with $xyz$ symmetry 
jumps at $T_p$, but this does not change the $T_d$ point-group 
symmetry.

The main purpose of this paper is to clarify the properties of 
anharmonic oscillations in tetrahedral symmetry, 
which is the local
symmetry for $A$-cations in $A$Os$_2$O$_6$, and 
the superconductivity mediated by these 
 anharmonic oscillations. In order to fully take into account the 
anharmonicity and anisotropy, we will solve the three-dimensional Schr\"odinger 
equation for an anharmonic
oscillator in the tetrahedral symmetry. Using these exact phonon eigenstates,
 we will then discuss the strong coupling superconductivity assuming an
 $s$-wave pairing.

This paper is organized as follows. In Sec. \ref{anharmo}, we will calculate
the energy spectrum of the anharmonic potential problem and discuss 
various thermodynamic and dynamical quantities and their dependence on
temperature. 
Anharmonic effects on
Debye-Waller factor will also be discussed. Section \ref{SCS} is devoted
to the discussions for strong coupling superconductivity mediated by the
anharmonic ion oscillations discussed in Sec. \ref{anharmo}. In Sec. 
\ref{discuss}, we will discuss the relevance of the present results to
the $\beta$-pyrochlore compounds. We also apply our theory to discuss 
the changes in phonon dynamics at the first-order transition at $T_p$. 
Finally Sec. \ref{conclusion}
is a summary of this paper.

\section{Anharmonic phonons}\label{anharmo}

\subsection{Model}\label{model}
In this paper, we investigate an anharmonic oscillation of K ion in KOs$_2$O$_6$.
Our model is anharmonic local phonons
at each lattice point and we assume the local symmetry is tetrahedral
one which corresponds to the case of K-site symmetry in KOs$_2$O$_6$.
In tetrahedral symmetry, in addition to spherical and cubic fourth order terms, 
there exists a third-order anharmonic term which
breaks inversion symmetry and the Hamiltonian is given by

\begin{eqnarray}
H&=&-\frac{\hbar^2}{2M}\nabla^2_{\bf R}+V({\bf R}),\label{eqH}\\
V(\bf R)&=&\frac{1}{2}M\Omega^2_0|{\bf R}|^2
         +BXYZ\nonumber\\
&&+C_1|{\bf R}|^4+C_2 (X^4+Y^4+Z^4),
\label{eq:pot}
\end{eqnarray}
where ${\bf R}=(X,Y,Z)$ is the real-space displacement of the ion from
 the equilibrium position. $M$ is the mass of the ion. Throughout this 
paper we set $M/m_e=71748$, where $m_e$ is the mass of electron, except 
for discussions in Sec. \ref{DiscussChemi}, and this corresponds to the
 mass of K ion. $\Omega_0$ and $B$
 are coefficients of second- and third-order terms in the
 ion potential, while $C_1$ and $C_2$ are isotropic and cubic
 fourth-order 
terms, and we ignore the higher-order potential terms
 of $O(R^5)$. To study dependence on potential parameters, it 
is useful to rewrite Hamiltonian (\ref{eqH}) into a dimensionless
 form by renormalizing displacement and energy by their units. 
As for the energy unit, we choose the energy of harmonic phonon
 corresponding the second-order term $\hbar \Omega_0$. The unit of the
 length is chosen as 
$a_0 \equiv \hbar/M\Omega_0=a_B\sqrt{2(e^2/2a_B)(m_e/M)/\hbar
 \Omega_0}$. Here, $a_B$ is the Bohr
 radius $a_B= \hbar^2/m_ee^2\simeq 0.53$ \AA.

 In these units, Hamiltonian (\ref{eqH}) is transformed to
\begin{eqnarray}
\bar{H}&\equiv&\frac{H}{\hbar \Omega_0}=-\frac{1}{2}\nabla^2_{\bf
 r}+\bar{V}(\bf r),\label{H00}\\
\bar{V}(\bf r)&=&\frac{1}{2} \omega_0|{\bf r}|^2
         +bxyz+c_1|{\bf r}|^4+c_2 (x^4+y^4+z^4)\label{H},
\label{eq:potential}
\end{eqnarray}
where ${\bf r}=(x,y,z)={\bf R}/a_0$. $b$ and $c_1$ and $c_2$
 are dimensionless constants and $\omega_0$ is introduced to vary the
 second-order term for later purpose, but $\omega_0=1$ for the main part of
 this paper. In Secs. \ref{SCS} and \ref{discuss}, 
we will use different values of $\omega_0$ to examine the effects of
 changes in the second-order term.\cite{omega0var} 
Throughout this paper we set $\hbar\Omega_0=44$ K which
corresponds to the energy scale of the optical-phonon frequency observed in
$\beta$-pyrochlore compounds,\cite{SummaryHiroi,NeutronSasai, Mukta}
and thus $a_0\simeq a_B/\sqrt{10}$.

In the limit of $b=c_1=c_2=0$, Hamiltonian (\ref{H00}) can be
 diagonalized by using creation and annihilation operators as,
\begin{eqnarray}
\bar{H}&=&a_x^{\dagger}a_x+
a_y^{\dagger}a_y+a_z^{\dagger}a_z+\frac{3}{2},\label{H0}
\end{eqnarray}
where 
$a_{\mu}\equiv\frac{1}{\sqrt{2}}(\partial_{\mu}+ x_{\mu})$
with $\mu=x$, $y$ or $z$. Eigenstates are labeled by three occupation numbers
 as $\bar{H}|n_x,n_y,n_z\rangle=(n_x+n_y+n_z+3/2)|n_x,n_y,n_z\rangle$ and 
$a^{\dagger}_{\mu}a_{\mu}|n_x,n_y,n_z\rangle=n_{\mu}|n_x,n_y,n_z\rangle$.

For nonzero $b$, $c_1$, and $c_2$, we diagonalize Hamiltonian (\ref{H00}) numerically
 in the restricted Hilbert space spanned by $\{|n_x,n_y,n_z\rangle\}$ with 
$n_x+n_y+n_z\le n_{\rm max}$. In this paper, we use $n_{\rm max}=40$
 which corresponds to the Hilbert space with 12341 states and 
 check the convergence by comparing the results for
 $n_{\rm max}=50$ including 23426 states. 
We employ this approach rather than 
 the conventional self-consistent Gaussian
 approximation\cite{DahmUeda,Yamakage}. This is because the conventional 
 approximation does not work for the potential, Eq. (\ref{H}) 
since the third-order term cannot be decoupled as in the 
fourth-order terms. It is also important that, as will be shown
 later, 
energy differences between adjacent eigenstate multiplets are not the same, 
and this property cannot be described by the self-consistent Gaussian
 approximation.


\subsection{Energy spectrum}\label{sec22}
\begin{figure}[tb]
\begin{center}
    \includegraphics[width=0.44\textwidth]{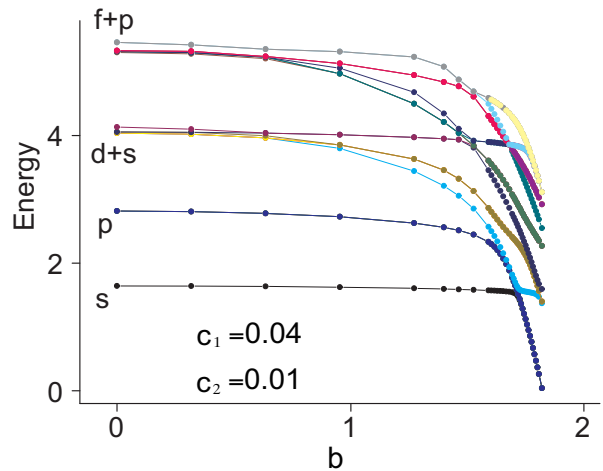}
\end{center}
\caption{(Color online) Energy eigenvalues vs the third-order anharmonic parameter 
$b$ for $c_1=0.04$ and $c_2=0.01$. $s$, $p$, $d$, and $f$ denote the approximate orbital
 symmetry, which is exact in the isotropic harmonic potential case.}
\label{fig-spec}
\end{figure}

In Fig. \ref{fig-spec}, we show a low-energy part of the energy spectra
of Hamiltonian
(\ref{H00}) as a function of the third-order anharmonicity $b$ for $c_1=0.04$ and $c_2=0.01$. 
The ground state is always nodeless in the space ${\bf r}$ and 
therefore singlet, 
while the first excited states are triplet, 
corresponding to $s$ and $p$ orbitals for the case of isotropic potential.
As $b$ increases, the energy of the lowest singlet excited state
(hereafter, this state will be referred to as
$s'$ state) decreases and shows an anticrossing with the ground state
 at $b=b^*\sim 1.7$. 
This kind of anti-crossing behavior does not occur in the 
one-dimensional anharmonic potential problem: $V(X)=aX^2+bX^4$.\cite{DahmUeda,Matsumoto}
Interestingly, near $b=b^*$ the five lowest-energy states are well
separated from other states in the energy spectra. Thus, the validity
of our previous five-state toy model is justified around this crossover region.\cite{Hat} 

\begin{figure}[tb]
\begin{center}
    \includegraphics[width=0.42\textwidth]{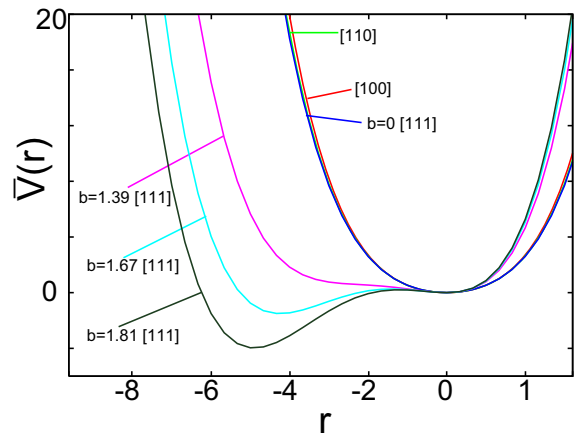}
\end{center}
\caption{(Color online) Potential along [100], [110], and [111]
 directions for various $b$. $c_1=0.04$ and $c_2=0.01$. ${\bf r}=r{\bf
 n}$, where $\bf n$ being the unit vector along each direction.}
\label{fig-pot}
\end{figure}

For $b>b^*$, the excited triplet states, which are $p$-wave like, are
nearly degenerate with the ground state. This means that these four
states are localized  away from the origin in the four directions: 
$[\bar{1}\bar{1}\bar{1}]$, $[11\bar{1}]$, $[\bar{1}11]$, and $[1\bar{1}1]$.
For the illustration, we show the potential form along
$[100]$, $[110]$, and $[111]$ directions in Fig.\ref{fig-pot}.
It is clear that an off-center potential local minimum emerges along
[111] direction as $b$ increases.
The energy of the first excited state from the ground state $\Delta$ is
small, owing to smallness of
 the quantum tunneling probability between different valleys. 
These four states form $sp^3$ orbitals and their energy can be well 
described by considering the quantum tunneling of the ion between the 
four potential minimum positions. Hereafter, we call these
states in this parameter region ($b>b^*$) as {\it quantum tunneling states}.


\subsection{Observables}
We now investigate the temperature dependence of the oscillation
amplitude of these anharmonic ions. 
Thermodynamic average of observable $A$ is calculated by the formula: 
\begin{eqnarray}
\langle A\rangle&=&\sum_{n}w_n \langle n| A |n\rangle,\label{Aav}
\end{eqnarray}
where $|n\rangle$ is the eigen state of Hamiltonian (\ref{eqH}) with
eigenenergy $E_n$ and $w_n$ is its Boltzmann weight $w_n=\exp(- E_n/T)/\sum_m \exp(-E_m/T)$,
 and $T$ is the temperature. 
In the summation in Eq. (\ref{Aav}), we discard states
with $E_n > 0.8n_{\rm max}\omega_0$, since the high-energy parts of our energy
spectrum is not correct due to the cutoff $n_{\rm max}$ and 
the high-energy part does not matter as long as the
low-temperature properties of the system are concerned.

\begin{figure}[tb]
\begin{center}
    \includegraphics[width=0.4\textwidth]{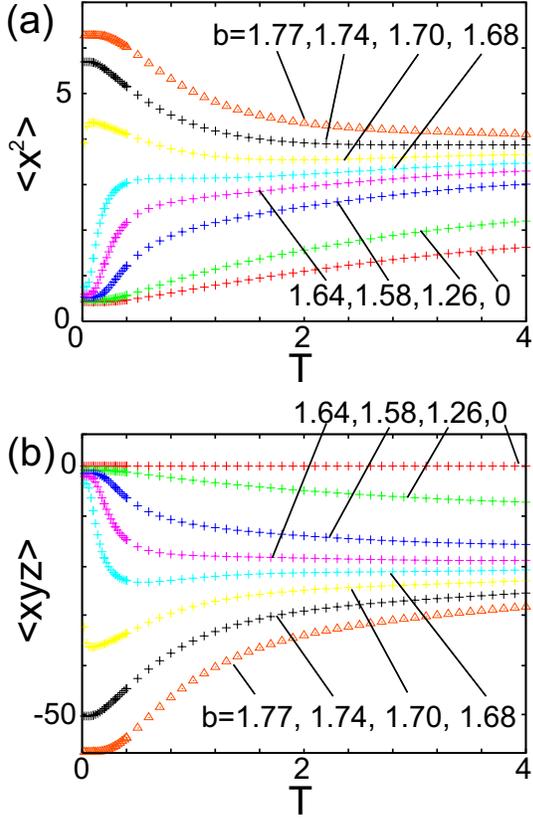}
\end{center}
\caption{(Color online) Temperature dependence of thermodynamic average (a) $\langle
 x^2\rangle$ and (b) $\langle xyz \rangle$. $c_1=0.04$ and $c_2=0.01$. }
\label{fig-x2xyz}
\end{figure}

In Fig \ref{fig-x2xyz}, we show the temperature
dependence of fluctuations $\langle x^2\rangle=\langle {\bf
r}^2\rangle/3$ and $\langle xyz\rangle$. 
Note that since 
$xyz$ is invariant quantity in tetrahedral symmetry, this does
not vanish except for $b=0$. As the temperature decreases, both $\langle
x^2\rangle$ and $|\langle xyz\rangle|$ decrease and saturate to finite
values for $b<b^*$. On the other hand, for $b>b^*$, these quantities
increase at low temperatures suggesting the quantum tunneling state.

\begin{figure}[tb]
\begin{center}
 \includegraphics[width=0.35\textwidth]{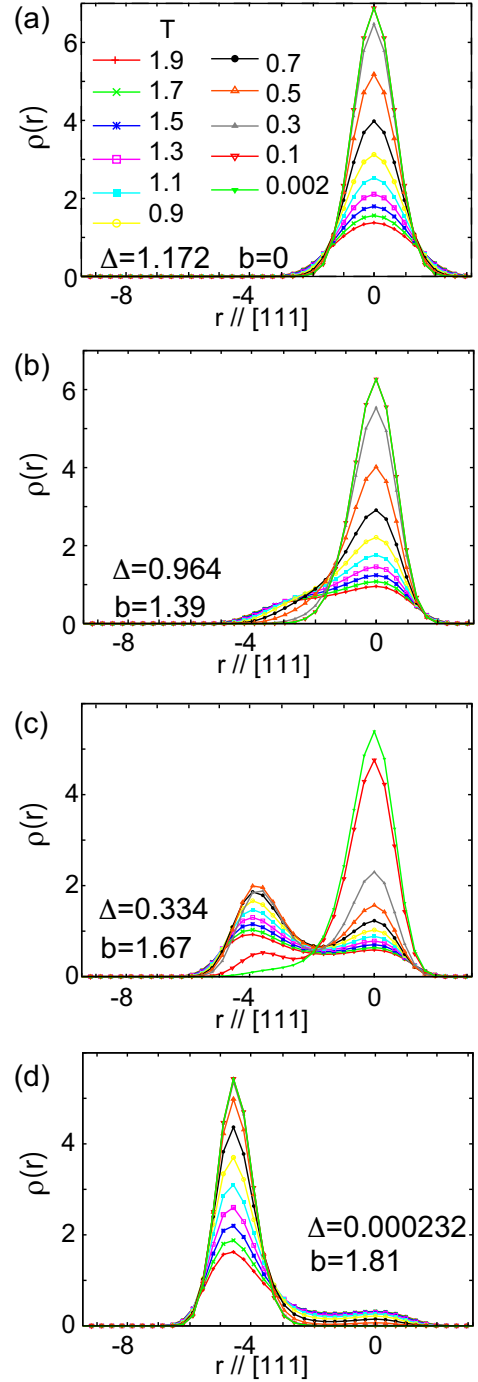}
\end{center}
\caption{(Color online) Atomic density profile $\rho({\bf r})$ along ${\bf r}=r(1,1,1)/\sqrt{3}$ for various
 $b$. $c_1=0.04$ and $c_2=0.01$.}
\label{fig-Dens}
\end{figure}

To see the change with $b$ more clearly, we calculate the 
ion density $\rho({\bf r})\equiv \sum_nw_n|\Psi_n({\bf
r})|^2$ at position ${\bf r}$. Here ${\Psi_n({\bf r})}$ is the
wave function at position ${\bf r} $ which is normalized 
as $\int d^3 x |\Psi_n({\bf r})|^2=1$.
In Figs. \ref{fig-Dens}(a)-\ref{fig-Dens}(d), we show $\rho(\bf r)$ along [111]
direction, ${\bf r}=r(1,1,1)/\sqrt{3}$, for several temperatures. 
$\rho({\bf r})$ in other directions
has monotonic temperature dependence and is peaked at 
${\bf r}=(0,0,0)$ as in  Fig. \ref{fig-Dens}(a). 
As $b$ increases, a part of the weight of $\rho(\bf r)$ 
shifts to the position around $r\sim -4$. Interestingly near $b=b^*$
[Fig. \ref{fig-Dens}(c)],
the temperature dependence of $\rho(\bf r)$ at $r\sim -4$, is
non-monotonic, 
{\it i.e.,} as $T$ decreases, 
$\rho(-4)$ increases first, but decreases below $T\sim 0.3\sim \Delta$. 
In Fig. \ref{fig-Dens}(d), the $\rho(0)$
is suppressed at low temperatures reflecting the nature of the quantum
tunneling states, since the potential minimum away from
the origin is deep enough as shown in Fig. \ref{fig-pot}.

\subsection{Debye-Waller factor}\label{debye-waller}
Now we discuss the Debye-Waller factor, which contains information about
the amplitude of ion oscillation and it is observed by elastic 
neutron-scattering experiments.\cite{NeutronText} Consider an atom 
 whose position is represented by  $\bf d$ in a unit cell
 and denote its displacement by ${\bf u_d}=(u_{\bf d}^x,u_{\bf
 d}^y,u_{\bf d}^z)$. 
Then, the Debye-Waller factor at the scattering wave vector ${\bf Q}$ 
 for this atom 
$e^{-W_{\bf d}({\bf Q})}$ is given by
\begin{eqnarray}
\exp[-W_{\bf d}({\bf Q})]= \langle \exp(-i{\bf Q \cdot u_d})\rangle. \label{DW0}
\end{eqnarray}
In the harmonic approximation, Eq. (\ref{DW0}) reduces to
\begin{eqnarray}
W_{\bf d}^{(2)}({\bf Q})=\frac{1}{2}\langle ({\bf Q \cdot u_d})^2\rangle. \label{DW1}
\end{eqnarray}
 Using $W_{\bf d}(\bf Q)$, the static structure factor $F(\bf
Q)$ is represented as
\begin{eqnarray}
F({\bf Q})&=&\sum_{\bf d}\bar{b}_{\bf d} \exp(i{\bf Q\cdot d}) \exp[-W_{\bf d}({\bf Q})],
\end{eqnarray}
where $\bar{b}_{\bf d}$ is the averaged scattering length of the atom 
at position ${\bf d}$.

When anharmonicity is not negligible, Eq. (\ref{DW1}) is not sufficient and
we must take into account higher-order terms. Up to the third order
in $(\bf Q\cdot u_d)$, we obtain
\begin{eqnarray}
W^{(3)}_{\bf d}({\bf Q})=\frac{1}{2}\langle ({\bf Q \cdot u_d})^2\rangle
+\frac{i}{6}\langle ({\bf Q \cdot u_d})^3\rangle.
\end{eqnarray}
As an example, let us consider K atoms in
KOs$_2$O$_6$, which constitute a diamond sub-lattice. 
 There are two sites in the unit cell, ${\bf d}_A=(0,0,0)$
and ${\bf d}_B=\frac{a}{4}(1,1,1)$. Here $a$ is the
lattice constant. From the symmetry arguments, we obtain
\begin{eqnarray}
\frac{1}{2}\langle ({\bf Q \cdot
  u_d})^2\rangle&=&\frac{1}{6}|{\bf Q}|^2 \langle |{\bf
  u_d}|^2\rangle\equiv W'_{\bf d}, \label{W1}\\
\frac{1}{6}\langle ({\bf Q \cdot
  u_d})^3\rangle&=&Q_xQ_yQ_z\langle u_{\bf d}^xu_{\bf
  d}^yu_{\bf d}^z\rangle \equiv W''_{\bf d}. \label{W2}
\end{eqnarray}
In the diamond lattice structure, $W'_{{\bf d}_A}({\bf Q})=W'_{{\bf d}_B}({\bf Q})\equiv W'({\bf
 Q})$ and $W''_{{\bf d}_A}({\bf Q})=-W''_{{\bf d}_B}({\bf Q})\equiv
 W''({\bf Q})$.
Using Eqs. (\ref{W1}) and (\ref{W2}) and setting $\bar{b}\equiv \bar{b}_{{\bf d}_A}=\bar{b}_{{\bf d}_B}$, we obtain
\begin{eqnarray}
|F({\bf Q})|^2&=& 2\bar{b}^2 \exp[-2W'({\bf Q})]\nonumber\\
&\times&\Bigg\{1+{\rm cos}\Big[{\bf Q}\cdot ({\bf d}_A-{\bf d}_B) -2W''({\bf Q})  \Big]  \Bigg\}.\label{eq-FQ}
\end{eqnarray}
Thus, the effective $W({\bf Q})$ becomes
\begin{eqnarray}
W_{\rm eff}({\bf Q}) &=& W'({\bf
 Q})-\frac{1}{2}\log2\nonumber\\
&-& \log \Big|{\rm cos}\Big[\frac{1}{2}{\bf Q}\cdot ({\bf d}_A-{\bf d}_B) -W''({\bf Q})  \Big]  \Big|.\nonumber\\
\end{eqnarray}
It is important to note that since the $W''({\bf Q})$ depends on the
wave vector as $Q_xQ_yQ_z$, the effect of anharmonicity is anisotropic.
For example, at ${\bf Q}_{111}=(1,1,1)2\pi/a$, $W_{\rm eff}$ is given by
\begin{eqnarray}
W_{\rm eff}({\bf Q}_{111}) &=& W_{\rm eff}(-{\bf Q}_{111}),\nonumber\\
&=& W'({\bf
 Q}_{111})-\frac{1}{2}\log \Bigg\{1+{\rm sin}\Big[2W''({\bf
 Q}_{111})  \Big]  \Bigg\},\nonumber\\
\label{W111}
\end{eqnarray}
whereas at ${\bf Q}_{200}=(2,0,0)2\pi/a$, we obtain
\begin{eqnarray}
W_{\rm eff}({\bf Q}_{200})=W'({\bf Q}_{200})=\frac{4}{3}W'({\bf Q}_{111}).
\end{eqnarray}

The third-order contribution 
$\langle u_{\bf d}^xu_{\bf d}^yu_{\bf d}^z \rangle$ 
also modifies the extinction rule of the structure factor. 
According to Eq. (\ref{eq-FQ}), $|F({\bf Q})|^2=0$ at ${\bf Q}=(2,0,0)2\pi/a$
and also at $(2,2,2)2\pi/a$ 
due to the interference factor $1+\cos[{\bf Q}\cdot ({\bf d_A-d_B})]$ 
when $W''({\bf Q})$ is set to be zero. For $b\ne 0$,  
$\langle u_{\bf d}^xu_{\bf d}^yu_{\bf d}^z \rangle$ is finite and 
this leads to nonvanishing $W''({\bf Q})$ at ${\bf Q}=(2,2,2)2\pi/a$ 
but $W''({\bf Q})=0$ at ${\bf Q}=(200)2\pi/a$, since $W''({\bf Q})$ 
is proportional to $Q_xQ_yQ_z$. Therefore, $|F({\bf Q})|^2$ is
nonvanishing at ${\bf Q}=(2,2,2)2\pi/a$ while remains zero 
at ${\bf Q}=(2,0,0)2\pi/a$.

\begin{figure}[tb]
\begin{center}
    \includegraphics[width=0.4\textwidth]{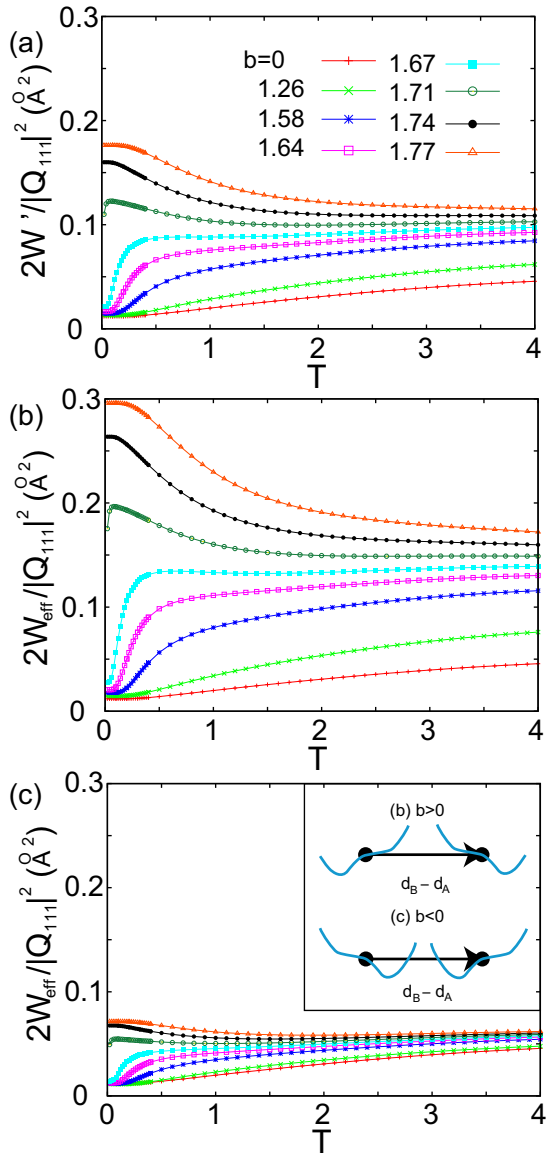}
\end{center}
\caption{(Color online) Temperature dependence of the exponent of the Debye-Waller factor for $c_1=0.04$
 and $c_2=0.01$. (a) $W'({\bf
 Q}_{111})$. (b) $W_{\rm eff}({\bf
 Q}_{111})$ for $b>0$. 
(c) $W_{\rm eff}({\bf
 Q}_{111})$ for $b<0$. 
Inset in (c): schematic potential form along the [111] direction in (b)
 and (c) for large $b$ cases.}
\label{fig-DebyeWaller}
\end{figure}

We show the calculated $W_{\rm eff}({\bf Q}_{111})$ for several values
of $b$ as a function of temperature in Fig. \ref{fig-DebyeWaller}. 
In Fig. \ref{fig-DebyeWaller}(a), the first term in Eq. (\ref{W111}) is
shown and $W_{\rm eff}({\bf Q}_{111})$ for $b>0$ and $b<0$ are shown in
(b) and (c), respectively. It is easily seen that, 
for small $b$, $W_{\rm eff}({\bf Q}_{111})$ in 
Figs. \ref{fig-DebyeWaller}(b) and \ref{fig-DebyeWaller}(c) are
essentially the same as $W'({\bf Q}_{111})$ in (a), while, for larger
$b$, $W_{\rm eff}({\bf Q}_{111})$ is enhanced (suppressed) for $b>0$ in
(b) [$b<0$ in (c)].  For other directions of ${\bf
Q}$ with $Q_xQ_yQ_z=0$, $W_{\rm eff}({\bf Q})$ does not include the
contribution of the third order term $W''(\bf Q)$ and thus, $W_{\rm eff}=W'$.

\begin{figure}[tb]
\begin{center}
    \includegraphics[width=0.5\textwidth]{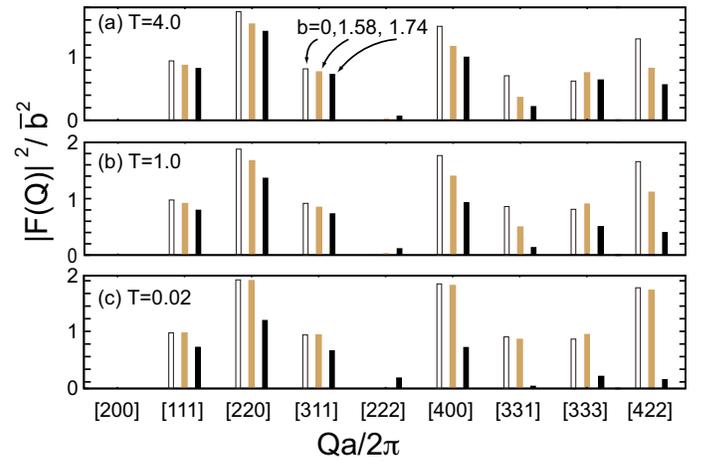}
\end{center}
\caption{(Color online) ${\bf Q}$ dependence of $|F({\bf Q})|^2$ for $c_1=0.04$
 and $c_2=0.01$ at temperatures (a) $T=4.0$, (b) $T=1.0$, and (c)
 $T=0.02$. 
For each of ${\bf Q}$, $|F({\bf Q})|^2$ values for $b=0$, $1.58$, and $1.74$ are shown from left
 to right. For ${\bf Q}=2\pi(2,0,0)/a$, $|F({\bf Q})|^2$ is always zero.}
\label{fig-DWQdep}
\end{figure}

Figure
\ref{fig-DWQdep} shows the wave-vector dependence of $|F({\bf
Q})|^2$ for several temperatures and $b$. Note that $F(\bf Q)$ at ${\bf
Q}=(2,0,0)2\pi/a$ vanishes due to the extinction rule mentioned above,
 while the intensity at ${\bf Q}=(2,2,2)2\pi/a$ appears for nonzero $b$
 although its strength is weak. Concerning $b$ dependence, 
$|F({\bf Q})|^2$ for $b=1.74$ is much
 suppressed for large $|{\bf Q}|$ and low temperatures, which is
 originated from the large displacement for $b>b^*$. As for the temperature dependence, $|F({\bf Q})|^2$
increases as $T$ decreases for $b=0$ and $1.58$, while 
 decreases for $b=1.74>b^*$. This
temperature dependence of $|F({\bf Q})|^2$ for $b>b^*$ is also
understood from the temperature dependence of $W_{\rm eff}$ 
in Fig. \ref{fig-DebyeWaller}, where $W_{\rm eff}$ 
for larger $b>b^*$ increases as $T$ decreases. 

We note that by careful analysis of the neutron data, as we have
demonstrated the effects of the third-order term on the Debye-Waller
factor, 
we can obtain useful information about anharmonicity in the real materials. 
In KOs$_2$O$_6$, since there are a lot of kinds of atoms such as the oxygen
and osmium, our results presented in this section cannot
 directly be compared with the neutron data of KOs$_2$O$_6$. It is
 desired to carry out high resolution neutron scattering experiments
 using the single crystal and analyze the results by taking into account the
 anharmonicity.


\subsection{Ion dynamics}\label{correlationfunc}
\begin{figure}[tb]
\begin{center}
    \includegraphics[width=0.48\textwidth]{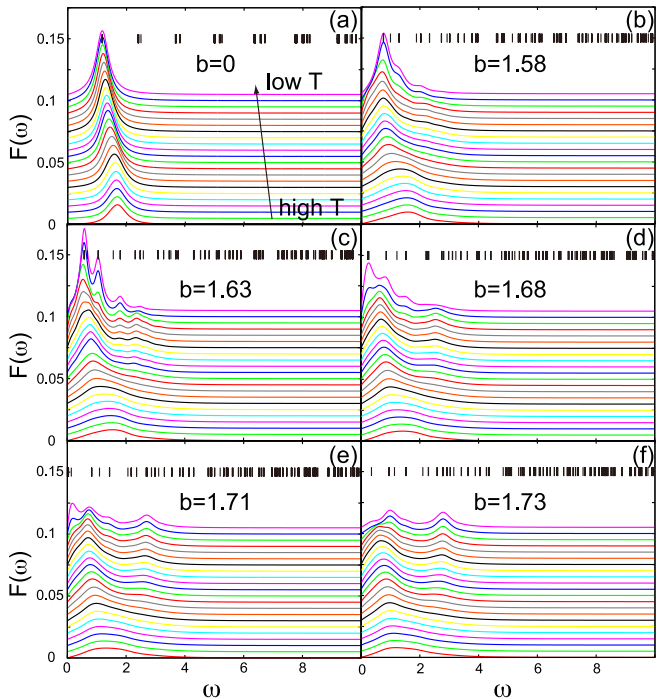}
\end{center}
\caption{(Color online) Phonon DOS $F(\omega)$ for various $b$'s
 fixing $c_1=0.04$ and
 $c_2=0.01$.  The data correspond to $T=$10,
 9.2, 8.4, 7.6, 6.8, 6.0, 5.2, 4.4, 3.6, 2.8, 2.0, 1.8, 1.6, 1.4, 1.2,
 1.0, 0.8, 0.6, 0.4, 0.2, and 0.1, from bottom to top and each curve is
 shifted up by 0.005. Energy
 eigenvalues relative to the energy of the ground state are 
shown by short bars in each panel.}
\label{fig-Spectralfunc}
\end{figure}

Now let us investigate the dynamics of the anharmonic ion oscillation.
In Fig. \ref{fig-Spectralfunc}, we show the phonon density of states
(DOS) $F(\omega)$, which is calculated from the 
correlation function $D$, 
\begin{eqnarray}
D_{\mu\mu'}(\omega)&=&\sum_{nm}(w_n-w_m) \frac{\langle n|x_{\mu}|m\rangle\langle m|x_{\mu'}| n \rangle}{\omega-E_m+E_n+i\eta}.\label{Dyn}
\end{eqnarray}
The DOS is given by its imaginary part, $F(\omega)=-{\rm Im}D_{xx}(\omega)/\pi$.
 Notice $D_{xx}(\omega)=D_{yy}(\omega)=D_{zz}(\omega)$ in the
 tetrahedral symmetry.
We set the phenomenological broadening parameter $\eta=0.26$. 
Although one can try more sophisticated analyses by 
introducing dissipation by electron-phonon
 couplings or couplings with other degrees of freedom to 
calculate  the
 imaginary part of the self-energy $\eta(\omega)$,\cite{Hat2,Takechi}
it is sufficient at this stage to restrict ourselves in
 the phenomenological level as long as characteristic properties of
 the DOS are concerned.

Figure \ref{fig-Spectralfunc}(a) shows the phonon DOS $F(\omega)$ 
 for the case $b=0$ and the eigenenergies relative to the ground
 state energy indicated by small bars. It is clearly seen that 
the position of the peak decreases
as the temperature decreases and this softening is consistent with the
 study based on self-consistent Gaussian approximations.\cite{DahmUeda,Yamakage} 
The single peak in $F(\omega)$ is actually constituted of many
 Lorentzian peaks each of which has a width of $\eta=0.26$. 
In this case of $b=0$, the transition matrix elements are sizable only between 
neighboring ``multiplets'' of the corresponding harmonic system. 
Because of the fourth-order anharmonicity, $c_1$ and $c_2$, 
the degeneracy of each multiplet is only approximate and 
slightly lifted, and also the energy separation between multiplets 
increases with increasing energy.
Since the distribution of Lorentzian peak positions is quite 
smooth and the energy dependence of Boltzmann weight 
is monotonic, they finally form a single broad peak. 
Its position shifts from a low energy at low $T$ where only the 
ground state is an important initial state to a higher energy at higher
 $T$ where higher excited states contribute more importantly.

For $b>0$, qualitatively different behaviors appear as shown in
Figs. \ref{fig-Spectralfunc}(b)-\ref{fig-Spectralfunc}(f). 
The transition matrix elements are now finite for many more 
pairs of eigenstates $\langle n|x|m\rangle$ for nonzero $b$. 
This is because the point-group symmetry 
changes from cubic $O_h$ to tetrahedral $T_d$ and 
the eigenstates 
$|n\rangle$ do not have definite parity related to inversion symmetry 
${\bf r}\to -{\bf r}$. 
In this point-group symmetry, for example, 
$x$ and $yz$ are bases of  the same $T_{2}$ irreducible 
representation. As a result, the matrix element such as 
$\langle yz|x|s\rangle$ is nonvanishing 
(here ``$s$'' means an isotropic $s$-wave state, {\it i.e.,} $A_{1}$ in
$T_d$ point group). 
In particular, the ground state has transition matrix elements much
larger than in the case of $b=0$.
Thus, even at the lowest temperature, more than one peaks are present 
in $F(\omega)$ for $b>0$, while the $b=0$ case has no 
additional visible high-energy peak in $F(\omega)$ as shown in
Fig. \ref{fig-Spectralfunc}(a), since 
the ground state has  nonvanishing matrix elements 
only with the odd-parity excited states and their 
magnitude is small.

For clarity, let us concentrate on a few specific peaks to understand 
their origin for $b\ne 0$.
 $F(\omega)$ for the lowest temperature $T=0.1$
in Fig. \ref{fig-Spectralfunc}(c) shows four peaks 
below $\omega\sim 3$. As it is easily checked by comparing the energy
eigenvalues indicated by bars in Fig. \ref{fig-Spectralfunc}, 
the positions of these peaks correspond to the excitation energies. 
Actually, 
the lowest three peaks in Fig. \ref{fig-Spectralfunc}
 (c) correspond to the transitions to the $T_2$ states which can be
 traced back to the $p$-wave sates in the one-phonon multiplet, the $d$-wave
 states in the two-phonon multiplet, and the $p$- or $f$-wave states in the
 three-phonon multiplet in the harmonic oscillator (see
 Fig. \ref{fig-spec} where the label of the multiplets in the harmonic
 oscillator are indicated). 

There are also peaks or shoulder-like structures
 originating from the transition between the $s'$ and $T_2$ states for $b>0$.
 Among them, the transition between the first excited $p$-wave states to $s'$
 state is clearly seen in Figs. \ref{fig-Spectralfunc}(b) and
 \ref{fig-Spectralfunc}(c). 
Since the energy difference between these excited states $\Delta'$ is 
smaller than that between the ground state and the first excited state 
$\Delta$, these
 shoulder-like structures appear at energies lower than the main peak
 position  which corresponds to the transition from the ground 
state and the
 first excited states for the intermediate temperatures. In
 Fig. \ref{fig-Spectralfunc}(d), $\Delta'$ is so small that the
 structure at $\omega\sim \Delta'$ is not clearly visible but the slope
 $F(\omega)/\omega$ near $\omega\sim 0$ is enhanced. 

In Figs. \ref{fig-Spectralfunc}(e) and \ref{fig-Spectralfunc}(f), 
the first excited states
are almost degenerate with the ground state, so that the lowest energy
peak in the figures does not correspond to the transition to the first
excited states but that between the first excited states and $s'$ state.

\section{Strong coupling superconductivity}\label{SCS}
In this section, we will discuss the strong coupling theory of superconductivity\cite{Eliashberg,Scalapino}
 in which the attractive force between electrons is mediated by the
 anharmonic phonons discussed in the previous section. 
The main issue of this section is the effect of the phonon anharmonicity
 on the superconducting transition temperature $T_{\rm c}$. 

We assume that the phonon DOS is 
 given by $F(\omega)$ obtained in Sec. \ref{correlationfunc}, {\it i.e.,}
 we ignore the renormalization due to the electron-phonon
 coupling. Generally, electron-phonon couplings lead to frequency 
renormalization and a finite life time of phonons, both of which are described in 
the renormalization of $F(\omega)$. This renormalization is 
 interesting and generally should be included, but the
 present calculation reproduces correct qualitative behaviors and 
we leave full self-consistent calculation  as a future problem. 
It is noted that the frequency renormalization is taken into account such that the 
potential parameters are chosen to reproduce a renormalized frequency.
The superconducting transition temperature $T_{\rm c}$ is
 obtained by applying the
 conventional strong coupling theory of $s$-wave pairing\cite{Eliashberg,Scalapino} to our
 anharmonic phonon system coupled to isotropic electron gas. 

Throughout this section, we use Matsubara formalism, which is efficient to
 determine $T_{\rm c}$. Matsubara
 formulation has also advantage that the 
 phenomenological broadening factor $\eta$ used in Sec.
 \ref{correlationfunc} is not necessary.

\subsection{Gap equations}
Following the conventional theory of strong coupling 
superconductivity,\cite{AGD} the transition
temperature $T_{\rm c}$ for isotropic s-wave gap $\Delta_{\rm SC}$ is determined by
the gap equation,
\begin{eqnarray}
\Delta_{\rm SC}(i\epsilon_m) &=&-\alpha^2 T\sum_n
 {
 K}(i\epsilon_n)D(i\epsilon_m-i\epsilon_n)\Delta_{\rm SC}(i\epsilon_n),\ \ \ \ \  \\
{ K}(i\epsilon_n)&=&\int_{-\infty}^{\infty}d\xi
 G(\xi,i\epsilon_n)G(\xi,-i\epsilon_n),
\end{eqnarray}
where the electron Green's function is given by
\begin{eqnarray}
G(\xi,i\epsilon_n)=\frac{1}{i\epsilon_n-\xi-\Sigma(i\epsilon_n)},
\end{eqnarray}
and the normal self-energy is given by
\begin{eqnarray}
\Sigma(i\epsilon_n)&=& -\alpha^2T\sum_m\int_{-\infty}^{\infty} d\xi
G(\xi,i\epsilon_n-i\nu_m)D(i\nu_m).\ \ \ \ \ \label{sig}
\end{eqnarray}
Here $D(i\nu_m)=D_{\mu\mu}(i\nu_m)$ is the phonon Green's function. 
$\epsilon_n=(2n+1)\pi T$ and
$\nu_m=2m\pi T$ are the fermionic and bosonic Matsubara frequencies, respectively. 
$\alpha^2$ is proportional to the square of the electron-phonon coupling
constant times electron DOS at Fermi energy. In
our model, it is a frequency-independent quantity and is set as
$\alpha^2=595$ K/\AA$^2$, which corresponds to $0.38$ in our units of energy and length. As in the conventional
theory\cite{Eliashberg,Scalapino}, the normal self-energy, Eq. (\ref{sig}) is essentially given by
the second-order perturbation theory and $K(i\epsilon_n)$ is
analytically given by\cite{Bergmann}
\begin{eqnarray}
K(i\epsilon_n)=\Big|\frac{\epsilon_n}{\pi}-\alpha^2 T\Big[2\sum_{m=1}^n
D(i\nu_m)+D(0)\Big]\Big|^{-1}.
\end{eqnarray}

To determine $T_{\rm c}$, we define
\begin{eqnarray}
\mathcal{M}_{mn}&\equiv&
-\alpha^2T\sqrt{{K}(i\epsilon_m)}D(i\epsilon_m-i\epsilon_n)\sqrt{{K}(i\epsilon_n)},\
\ \ \ \ \\
&=&\mathcal{M}_{nm},\\
\psi_n&\equiv&
\Delta_{\rm SC}(i\epsilon_n)\sqrt{{K}(i\epsilon_n)},
\end{eqnarray}
 and solve eigenvalue problem numerically
\begin{eqnarray}
\Lambda(T)\psi_m &=&\sum_n {\mathcal M}_{mn}
 \psi_n.\label{eigeneq}
\end{eqnarray}
With decreasing temperature, $\Lambda(T)$ increases and $T_{\rm c}$ 
is given by the condition: $\Lambda(T_{\rm c})=1$.

\subsection{Superconducting transition temperature in harmonic systems}
\begin{figure}[tb]
\begin{center}
    \includegraphics[width=0.44\textwidth]{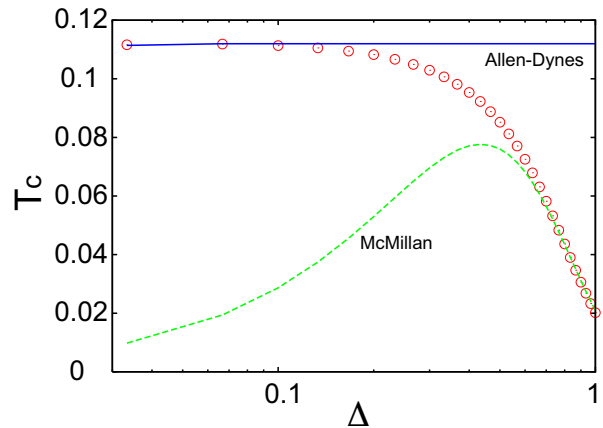}
\end{center}
\caption{(Color online) Superconducting transition temperature $T_{\rm c}$ vs the
 phonon energy $\Delta$ for the harmonic case $b=c_1=c_2=0$. The dashed line
 corresponds to $T^M_{\rm c}$ given by Eq. (\ref{McMillan}) and the solid line
 represents $T_{\rm c}^{AD}$ given by Eq. (\ref{allen})} 
\label{fig-Tscharmo}
\end{figure}

Before discussing the effects of phonon anharmonicity on 
 $T_{\rm c}$, we show in Fig. \ref{fig-Tscharmo} $T_{\rm c}$ in the harmonic
 system as a function of the first excited state energy
$\Delta=\omega_0$. The dash line is $T_{\rm c}$ calculated by 
McMillan's formula\cite{McMillan}, which is valid in the weak-coupling
regime, while the solid line is that by Allen and
Dynes,\cite{AllenDynes} which gives a good estimation of $T_{\rm c}$ in the
extremely strong coupling regime.  Let us summarize two analytic
formulae for the two opposite limits. 
The McMillan's formula is given by
\begin{eqnarray}
T_{\rm c}^{M}&=& 0.8\omega_{\rm log}\exp\Big(-\frac{1+\lambda}{\lambda}\Big),\label{McMillan}
\end{eqnarray}
where the Coulomb pseudo-potential
 is not taken into account. Here, the dimensionless
coupling constant $\lambda$ is given by 
\begin{eqnarray}
\lambda\equiv\alpha^2\int_{-\infty}^{\infty}
 \!\!\!\!\!\!d\omega \ 
\frac{F(\omega)}{\omega}
= \alpha^2\sum_{n,m}
 \frac{w_n-w_m}{E_{m}-E_n}|\langle n|x|m\rangle|^2,
\end{eqnarray}
and the characteristic energy scale $\omega_{\rm log}$ is given by
\begin{eqnarray}
&&\log \omega_{\rm log} = \frac{2\alpha^2}{\lambda}\int_0^{\infty} d\omega
 \frac{F(\omega)}{\omega}  \log \omega, \\
&&=\frac{2\alpha^2}{\lambda}\sum_{E_m>E_n}
 \frac{w_n-w_m}{E_{m}-E_n}|\langle n|x|m\rangle|^2\log{(E_m-E_n)}.\ \ \
\end{eqnarray}
 The Allen-Dynes formula is given by
\begin{eqnarray}
T^{AD}_{\rm c}=0.18\sqrt{\lambda\langle \omega^2\rangle},\label{allen}
\end{eqnarray}
where $\langle \omega^2\rangle=2\alpha^2\int_0^{\infty} \omega
F(\omega)d\omega/\lambda$. 

In the case of harmonic phonons,
$\lambda$ can be analytically calculated and is given by
$\lambda=\alpha^2/\Delta^2$.\cite{dimension} This means the smaller $\Delta$ is, the
larger $\lambda$ is realized. 
The McMillan's formula (\ref{McMillan}) indeed reproduces 
$T_{\rm c}$ for large
 $\Delta$ as expected, while for small $\Delta$, $T_{\rm c}$
 approaches the value of the Allen-Dynes formula (\ref{allen}) 
($=0.18\sqrt{0.38}\simeq 0.11$) as shown
 in Fig. \ref{fig-Tscharmo}. Note that for the harmonic phonons, $\langle
 \omega^2 \rangle=\Delta^2$. This relation is valid only for harmonic
 oscillators as will be discussed below. 

\subsection{Superconducting transition temperature in anharmonic
  systems}\label{anharmoTsc}
\begin{figure}[tb]
\begin{center}
    \includegraphics[width=0.44\textwidth]{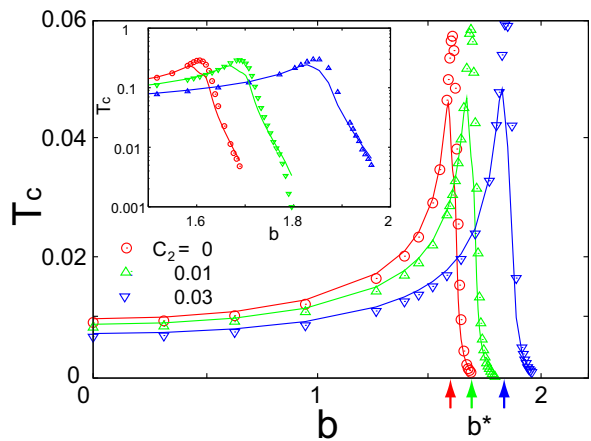}
\end{center}
\caption{(Color online) $T_{\rm c}$ vs $b$ for $c_1=0.04$ and $c_2=$0, 0.01, and
 0.03. The solid lines show $T_{\rm c}$'s of McMillan's formula
 (\ref{McMillan}). $b^*$'s for each of $c_2$ are indicated by arrows. 
Inset: zoom up in the large $b$ part.}
\label{fig-Tsc}
\end{figure}

Now, we discuss the effects of anharmonicity of phonon dynamics on $T_{\rm c}$.
Figure \ref{fig-Tsc} shows the dependence of $T_{\rm c}$ on the
 third-order anharmonic term $b$ of the ion potential, Eq. (\ref{H}) for
 $c_2=0$, $0.01$, and $0.03$, at fixed $c_1=0.04$. All the three cases exhibit a pronounced
 peak around the crossover value $b=b^*$ as discussed in Sec. \ref{anharmo}
 for each $c_2$, and $T_{\rm c}$'s are strongly suppressed 
in the quantum tunneling states, {\it i.e.,} for large $b>b^*$. 
The solid lines in the figure represent the McMillan's formula $T^M_{\rm
 c}$ [Eq.  (\ref{McMillan})]. The agreement with the McMillan's formula 
 indicates that $T_{\rm c}$ is
 qualitatively given by the conventional theory of strong coupling 
superconductivity. Note that $\lambda$ and
$\omega_{\rm log}$ vary with the temperature in the presence of
anharmonicity and $T_{\rm c}^M$ are calculated by using these values evaluated at $T=T_{\rm c}$.
For $b\simge b^*$, however, the 
discrepancy becomes large as shown in the inset of Fig. \ref{fig-Tsc}, and the effects of anharmonicity and the nature of the extreme
 strong coupling regime appear, which will be discussed later in Sec. \ref{exe}. 


\subsection{Competition between the energy scale and coupling constant}\label{coupl}
In Sec. \ref{coupl}, we discuss the origin of the peak structure based on the
McMillan formula.\cite{McMillan}

\begin{figure}[tb]
\begin{center}
    \includegraphics[width=0.44\textwidth]{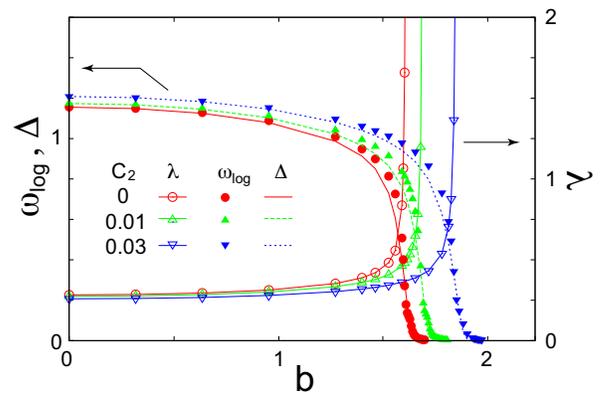}
\end{center}
\caption{(Color online) $\lambda$, $\omega_{\rm log}$, and $\Delta$ vs $b$ for $c_2=0$,
 $0.01$, and $0.03$. The open
 symbols represent $\lambda$ and the filled symbols with a solid line
 represent $\omega_{\rm log}$. Lines without symbols 
 represent $\Delta$. 
 $\lambda$ and $\omega_{\rm log}$ are calculated at $T=T_{\rm c}$.}
\label{fig-lam}
\end{figure}

The McMillan's formula is expressed with the two parameters, $\lambda$
and $\omega_{\rm log}$, as shown in Eq. (\ref{McMillan}). 
In Fig. \ref{fig-lam}, we show the $b$ dependence of $\lambda$ and
$\omega_{\rm log}$ calculated at $T_{\rm c}$ together with $\Delta$. 
It is noticeable that $\lambda$ 
shows a steep increase above $b^*$ and $\omega_{\rm log}$ is almost 
the same as $\Delta$ for $b\simle b^*$.
From these facts, one can easily understand that $T_{\rm c}$ calculated by
Eq. (\ref{McMillan}) decreases in the quantum tunneling states.
 The peak structure in $T_{\rm c}$ in
Fig. \ref{fig-Tsc} is a result of the competition between the
suppression of $\omega_{\rm log} (\Delta)$ and the enhancement of $\lambda$.
Thus, the peak structure is realized at $b\simeq b^*$, where
 $\lambda$ is large enough and simultaneously $\omega_{\rm log}$ is
not vanishingly small.

\subsection{Strong coupling limit}\label{exe}
In an extreme strong
coupling regime $\lambda \gg 1$, 
Allen and Dynes showed that  $T_{\rm c}$ is
approximately given by Eq. (\ref{allen}) rather than  the McMillan's
formula (\ref{McMillan}).\cite{AllenDynes}
 When Eq. (\ref{allen}) is applied to our local phonon
problem, we need care about the quantity $\lambda\langle
\omega^2\rangle$. As pointed out by Hardy and Flocken,\cite{Hardy} this quantity is related to
the f-sum rule and this implies $T_{\rm c}$ is universal 
 in any potential and 
depends only on the mass of the ion and $\alpha^2$ as shown in Fig. \ref{fig-Tscharmo}. However, 
this is not true in our anharmonic phonon system. From numerical calculations,
instead of $\langle \omega^2 \rangle$, we
find that $T_{\rm c}$ for $b\gg b^*$ is approximately given by
\begin{eqnarray}
T_{\rm c}=0.18\Delta\sqrt{\lambda}.\label{allen2}
\end{eqnarray}
We note that the origin of large $\lambda$ for $b>b^*$ is 
the strong suppression of $\Delta$ for
$b>b^*$ and other high-energy states do not play an important
role.  Thus, $\sqrt{\langle \omega^2 \rangle}$ overestimate the energy
scale of $T_{\rm c}$. These facts support the use of $\Delta$ instead of
the averaged frequency $\sqrt{\langle \omega^2 \rangle}$.

\begin{figure}[tb]
\begin{center}
    \includegraphics[width=0.44\textwidth]{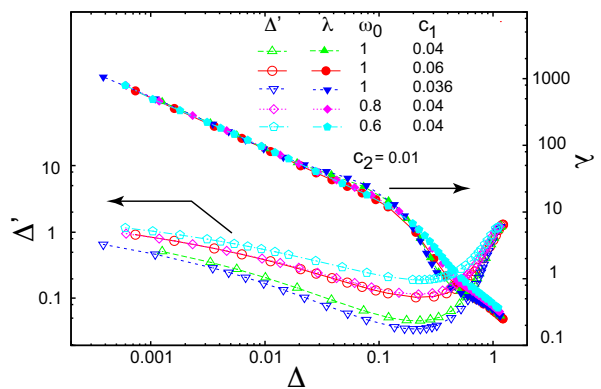}
\end{center}
\caption{(Color online) $\lambda$, $\Delta'$ vs $\Delta$ for $c_2=0.01$ and five 
 sets of $\omega_0$ and $c_1$. The open (filled) symbols represent
 $\Delta'$ ($\lambda$). $\Delta$ is controlled by
 varying $b$.}
\label{fig-lambda-Es-Delta}
\end{figure}

\begin{figure}[tb]
\begin{center}
    \includegraphics[width=0.44\textwidth]{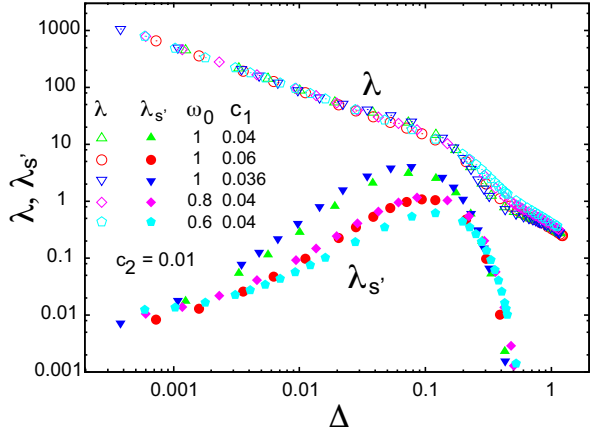}
\end{center}
\caption{(Color online) $\lambda$ and $\lambda_{s'}$ vs $\Delta$ for $c_2=0.01$ and
 five sets of $\omega_0$ and $c_1$. The open (filled) symbols
 represent $\lambda$ ($\lambda_{s'}$). }
\label{fig-lam_i}
\end{figure}

\subsection{Additional channel of attractive interactions}\label{additional}
As we noted before, there is a crossover at $b=b^*$ to the quantum
tunneling states and this means
there are five states at low-energy region of the spectra as shown in
Fig. \ref{fig-spec}. In the harmonic case, $\lambda$ is basically
determined by the matrix element $\langle n|x|0\rangle$ 
between the ground state $|0\rangle$ and the threefold-degenerate first excited states, and
the corresponding excitation energy $\Delta$. Since four low-energy
excited states are nearly degenerate with the ground state 
 around $b=b^*$, the singlet excited state $|s'\rangle$ also 
has a noticeable matrix element $\langle
n|x|s'\rangle$.
This provides another channel of 
attractive interaction and contributes to
$T_{\rm c}$. It is important to note that this
additional contribution never appears in the one-dimensional model and 
is a direct consequence of the crossover of the ground state at $b=b^*$.

In Fig. \ref{fig-lambda-Es-Delta}, we show 
$\lambda$ and the energy difference between $s'$ state and the $p$-wave like
first excited states: $\Delta'=E_{s'}-E_p$. 
They are calculated with varying $b$, and plotted as a function of the
 energy difference $\Delta$ between the first excited state and the ground
 state. 
$\lambda$ increases roughly as
$\sim \Delta^{-3/4}$ for small $\Delta$. $\Delta'$ shows minimum around
$\Delta\sim 0.2$ and this reflects the anti-crossing in the energy spectra.  
Corresponding to this, the coupling constant $\lambda_{s'}$ related
to $s'$ state shows a maximum around $\Delta\sim 0.1$ as shown in
Fig. \ref{fig-lam_i}. Here, we define $\lambda_{s'}$ as
\begin{eqnarray}
\lambda_{s'}=2\alpha^2\sum_{n=p}
 (w_n-w_{\rm s'})|\langle n|x|{\rm s'}\rangle|^2/(E_{\rm s'}-E_n),\label{lamdas}
\end{eqnarray}
where the $n$ summation is taken over the three first excited states.
 Note that the smaller $\Delta'$ is, the larger $\lambda_{s'}$ is
 obtained. Depending on the magnitude of $\lambda_{s'}$, total $\lambda$
 has a small bump 
 around the $\Delta\sim 0.1$ as a function of $\Delta$. This observation
 clearly shows that there is a new channel of interaction via the
 excitations between the first excited states and the $s'$ state.

\subsection{Deviations from Allen-Dynes formula near $b=b^*$}\label{devi}

As we discussed in Sec. \ref{additional}, the enhancement 
of $\lambda_{s'}$ influences the total $\lambda$ and 
 also affects $T_{\rm c}$ itself. 
In Sec. \ref{devi}, we derive a formula which includes the contribution
of $\lambda_{s'}$. 
In order to take into account this contribution in
Eq. (\ref{allen2}), we assume the DOS given by
\begin{eqnarray}
F(\omega)=\frac{\omega}{2}\Big[\lambda_p\delta(\omega-\Delta) + \lambda_{s'}\delta(\omega-\Delta')\Big]. \label{F12}
\end{eqnarray}
Since the enhancement of $\lambda_{s'}$ takes place in a finite but 
 extremely strong coupling regime,
 the total $\lambda$ is large and we follow the discussion by
 Allen and Dynes to estimate the lower bound of
$T_{\rm c}$, by assuming a trial gap function $\Delta_{\rm
SC}(i\omega_n)=\Delta_{\rm SC}^0\delta_{|\omega_n|,\pi T}$.\cite{AllenDynes} 
Using the form (\ref{F12}), we obtain the lower bound of $T_{\rm c}$ as
\begin{eqnarray}
8\pi^2T_{\rm c}^2&=&\lambda_p\Delta^2+(\lambda_{s'}-1){\Delta'}^2\nonumber\\
                &+&\sqrt{[\lambda_p\Delta^2+(\lambda_{s'}-1){\Delta'}^2]^2+4(\lambda_p+\lambda_{s'})\Delta^2{\Delta'}^2},\nonumber\\
\label{solution0}
\end{eqnarray}
In the limit of $\lambda_{s'}\to 0$ Eq. (\ref{solution0}) reduces 
to Eq. (\ref{allen2}): $T_{\rm c}=\Delta\sqrt{\lambda_p}/(2\pi)$.
Although the more sophisticated gap function can give the almost correct
coefficient $0.18$ instead of $(2\pi)^{-1}$,\cite{AllenDynes} it is
enough to consider 
 the simplest trial gap function in order to examine the contributions
of the $s'$ state to $T_{\rm c}$.
In Eq. (\ref{solution0}), we assume
$\lambda_p (\gg 1)$ corresponds to the coupling constant originating from the
transition between the ground state and $p$-wave like states (first excited states)
and $\lambda_{s'} \ll \lambda_p$ is the part from $s'$ state
[Eq. (\ref{lamdas})]. Since $\lambda_p\gg\lambda_{s'}$, we
can obtain a simpler approximate form of Eq. (\ref{solution0}). This
 is obtained by
simply replacing $\lambda$ in Eq. (\ref{allen2}) by $\lambda_{\rm eff}$, 
\begin{eqnarray}
\lambda_{\rm eff}=
 \lambda_p\Bigg[1+\frac{\lambda_{s'}}{1+2\lambda_p\Big(\frac{\Delta}{\Delta'}\Big)^2}\Bigg]. \label{solution2}
\end{eqnarray}

\begin{figure}[tb]
\begin{center}
    \includegraphics[width=0.5\textwidth]{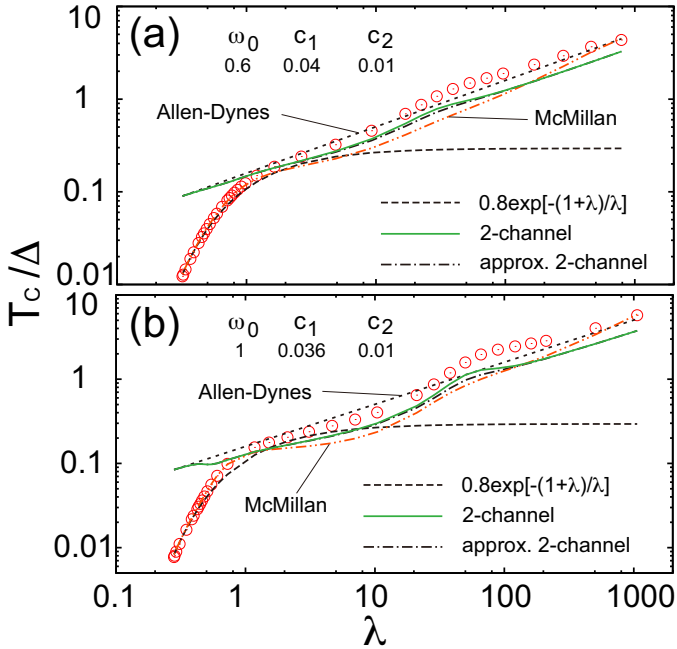}
\end{center}
\caption{(Color online) Various $T_{\rm c}$ formulas (\ref{McMillan}),
 (\ref{allen}), (\ref{solution0}), and (\ref{solution2}) vs
 $\lambda$. Formula (\ref{McMillan}) with $\omega_{\rm log}$ replaced by $\Delta$ is also
 plotted for comparison. (a) $\omega_0=0.6$, $c_1=0.04$, and $c_2=0.01$. (b)
 $\omega_0=1$, $c_1=0.036$, and $c_2=0.01$.}
\label{fig-scaling}
\end{figure}

Figure \ref{fig-scaling} shows  the $\lambda$ dependence of $T_{\rm
c}/\Delta$ for two typical sets of parameters in log-log scale. 
As shown with the dashed line, the Allen-Dynes formula shows a straight
line. For small $\lambda$, the calculated values agree with the
McMillan's formula, while for extremely large $\lambda$ they approach the line
of the Allen-Dynes formula. For the intermediate $\lambda>1$,
however, the results show a bending and this bending is larger in
Fig. \ref{fig-scaling}(b) than in Fig. \ref{fig-scaling}(a).
Our two-channel formula (\ref{solution0}) can describe the bending 
 as indicated by the solid line. Although the lower
bound of $T_{\rm c}$, Eq. (\ref{solution0}) is not in very good agreement with
the calculated $T_{\rm c}$, it
 captures the overall behavior of $T_{\rm c}$ for $\lambda>1$. It is
also noted that, although the McMillan's formula (\ref{McMillan}) shows
 an increase in $T_{\rm c}/\Delta$ for large $\lambda$, the
increase is too steep. The present two-component analysis confirms the
existence of the additional channel of the interaction near the
crossover point in this system.

\section{Discussions}\label{discuss}

Let us now discuss the implications of our calculations 
to compare with characteristic properties experimentally 
observed in $\beta$-pyrochlore compounds.  
One point is the chemical trends among 
the three member compounds, and we discuss why 
the potassium compound has the highest $T_{\rm c}$ of 
superconductivity.  Another point is about the question 
why the superconducting phase is not so much affected 
by the isomorphic structure transition at $T_p$ 
in the phase diagram. We will also propose a possible change in the
K-oscillation profile at $T_p$ and the effects of the transition on superconductivity.

\subsection{Chemical trends in $\beta$-pyrochlore compounds}\label{DiscussChemi}

Among the three $\beta$-pyrochlore compounds
$A$Os$_2$O$_6$ ($A$=K, Rb or Cs),  the K compound 
has the highest $T_{\rm c}^\mathrm{K}$=9.6 K of superconductivity 
and the strongest anharmonicity in the $A$-cation oscillation 
dynamics as observed in 
the neutron-scattering
experiments:
the Debye-Waller factor of K ion is much smaller than of  Rb and Cs
 compounds\cite{NeutronSasai} and the softening of the low-energy phonon peak 
 is also the strongest in KOs$_2$O$_6$.\cite{Mukta}
The Rb compound has the next 
strongest anharmonicity and the second highest $T_{\rm c}^\mathrm{Rb}$=6.3 K, 
while the Cs compound has the weakest anharmonicity 
and the lowest $T_{\rm c}^\mathrm{Cs}$=3.3 K.  
The ratio of $T_{\rm c}$ is approximately 3:2:1. 
Thus, the anharmonicity in the ion dynamics 
and the value of $T_{\rm c}$ are related. 
Let us examine this point in our results 
and also check if one can explain, at least qualitatively, 
the trends of other important quantities, 
 electron mass enhancement and phonon energy.  

A crucial difference among the three compounds is the 
size of $A$-cation; the K ion has the smallest size, followed 
by Rb, and Cs is the biggest ion.  
Since the size of the surrounding Os$_{12}$O$_{18}$ cage essentially does 
not change, the K ion has the largest space inside the cage,\cite{Yamaura}    
leading to strongly anharmonic oscillations. 
This reflects in different shapes of the $A$-cation potential, 
as shown in the calculation by Kune\v{s}, {\it et al.},\cite{BandCal} 
although the point-group symmetry is common.  

In our theory, the compound-dependent potential shapes are 
modeled by adjusting parameters in the potential, Eq. (\ref{eq:pot})
or (\ref{eq:potential}). 
In the following, we will show that $T_{\rm c}$ values and
the ion dynamics in the
three compounds are naturally explained by appropriate choice of potential 
parameters. Throughout Sec. \ref{DiscussChemi}, we will use 
some parameters explicitly shown with physical dimensions if necessary.

Let us first consider the trend of the value $b/b^*$ among the three compounds, 
recalling  the anharmonicity appearing in the Debye-Waller factor 
and the phonon energy observed in the neutron experiments.
\cite{NeutronSasai,Mukta}
 As shown in Fig. \ref{fig-Dens}, 
the anharmonicity grows with $b/b^*$. This implies that the $b/b^*$ 
value is the smallest for Cs, then for Rb, and the largest for K. 
Note that the largest ratio is still smaller than or at most equal to 
the crossover value $b/b^*=1$, since 
the observed K-cation density profile does not change 
its peak position from the equilibrium 
position.\cite{Yamaura,newSasai}

\begin{figure}[tb]
\begin{center}
    \includegraphics[width=0.5\textwidth]{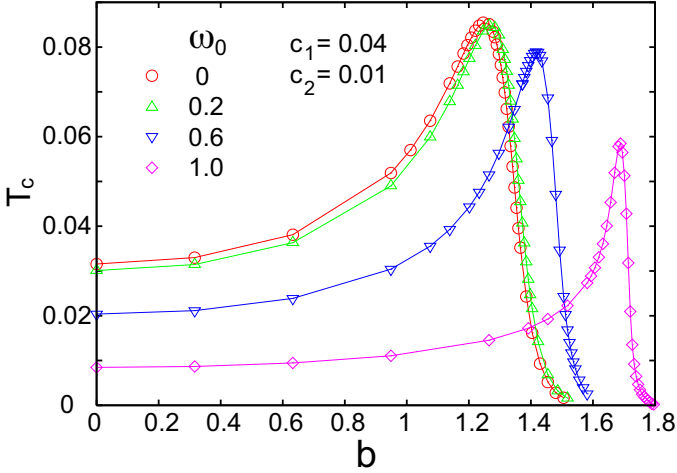}
\end{center}
\caption{(Color online) $T_{\rm c}$ vs $b$ for five values of $\omega_0$ fixing
 $c_1=0.04$ and $c_2=0.01$.} 
\label{fig-varOMEGA0}
\end{figure}

Secondly, we consider the trend of the second-order potential parameter 
$\omega_0$. Figure \ref{fig-varOMEGA0} shows
the $b$ dependence of $T_{\rm c}$ for several values of $\omega_0$.
The first-principle calculation of the ion potential clearly
 shows that $\omega_0$ is very small for K while larger for 
Rb and the largest for Cs.\cite{NoteBandCalc}
 This trend is consistent with the fact that the K compound has the strongest 
 anharmonicity and the highest $T_{\rm c}$ as far as $b<b^*$, as shown in
 Fig. \ref{fig-varOMEGA0}. It is also important to note that the peak
 position $\sim b^*$  becomes smaller as $\omega_0$ decreases.

By keeping these variations in $b/b^*$ and $\omega_0$ in mind, let us 
discuss our results of $\Delta$, $\lambda$ at $T=T_{\rm c}$ 
and $T_{\rm c}$ itself for each of
the three compounds. In order to make discussion simple, we assume
 the electron phonon coupling and the electron density of states are the 
 same for the three compounds and $\alpha^2=1897$ K/\AA$^2$. The atomic mass
 of each $A$-cation is $M_{\rm K}=71748m_e$, $M_{\rm Rb}=156839m_e$, and
 $M_{\rm Cs}=243890m_e$, respectively.
We have adjusted potential
 parameters, Eq. (\ref{eq:pot}), for each 
compound to reproduce the phonon
 energy and the effective ion oscillation variance $\langle x^2
 \rangle_{\rm exp}$ determined 
from the Debye-Waller factor.\cite{NeutronSasai,Mukta,Galati} 
As shown in Sec. \ref{debye-waller}, $W_{\bf d}({\bf Q})$ generally
contains the contribution of the third-order fluctuations $\langle
xyz\rangle$ in addition to $\langle x^2 \rangle$. However, since the 
experimental data is the ${\bf Q}$-averaged value of 
$\overline{ W_{{\bf d}}({\bf Q})/|{\bf Q}|^2}$, 
it is sufficient to consider the second-order fluctuations
$\langle W_{{\bf d}}({\bf Q})/|{\bf Q}|^2\rangle =\frac{1}{6}\langle |{\bf
 u_d}|^2\rangle=\frac{1}{2}\langle x^2 \rangle$.
 Since our potential, Eq. (\ref{eq:pot})
or (\ref{eq:potential}),  
includes four parameters, we fix $B=9324$ K/\AA$^3$ and
$C_1=4C_2=3332$ K/\AA$^4$ for simplicity and vary $\Omega_0$. It
is noted that $\Omega_0$ is  expected to become smaller as the size of the
alkali cation decreases,
and, for fixed $B$, $b/b^*$ becomes smaller as $\Omega_0$  
increases.  

Table \ref{tbl-1} shows the list of  
the basic quantities, $\Delta$, $\lambda$, and $T_{\rm c}$ for the 
three compounds. Let us first discuss $T_{\rm c}$. 
It is quite sure that one needs also to 
include a high-energy phonon, which plays a role to increase 
the $T_{\rm c}$ by about 3-5 K for all the three compounds, 
to obtain $T_{\rm c}$ consistent with the experimental results. 
$T_{\rm c}^{(2)}$ is one
 calculated with including an additional 
phonon with energy $\hbar\Omega_h=260$ K. 
We have set the corresponding dimensionless coupling constant
$\lambda_h=0.256$ in the calculations.
The results of 
$T_{\rm c}^{(2)}$ are quantitatively consistent with the experimental values $T_{\rm
c}^{\rm K}=9.6$ K, $T_{\rm c}^{\rm Rb}=6.3$ K, and $T_{\rm c}^{\rm Cs}=3.3$ K.
Here, we can reproduce the experimental values of $T_{\rm c}$ by
assuming the same $\alpha^2$, $\lambda_h$ and $\Omega_h$ for all the
three compounds and we conclude that 
the origin of the difference in $T_{\rm c}$ for the three
$\beta$-pyrochlore compounds is due to the difference mainly in the
anharmonicity of the alkali cation oscillations.\cite{InThisPoint}

Now, let us discuss the electron effective mass.
In our theory, the electron mass enhancement factor is related to 
$\lambda_{\rm tot}$ as $1/z_{\rm tot} = 1+\lambda_{\rm tot}$.
$\lambda$'s obtained from experiments and previous theoretical studies 
are $\lambda^{\rm K}=$1.6-2.4, \cite{SummaryHiroi,BattloggSC,Nagao,Chang} 
 for K, 
$\lambda^{\rm Rb}=$1.0-1.3, \cite{Nagao,Manalo} for Rb and
$\lambda^{\rm Cs}=0.78$,\cite{Nagao} for Cs, respectively. 
Our estimation of $\lambda_{\rm
tot}=\lambda+\lambda_h=\lambda+0.256$ with $\lambda$ in Table \ref{tbl-1} 
is qualitatively consistent with these values, although we have not 
fit $T_{\rm c}$ for each compound. 
The experimental values of specific-heat coefficient
$\gamma$ is $\gamma_{\rm K}=70$, $\gamma_{\rm Rb}=45$, and $\gamma_{\rm
Cs}=41$ mJ/mol K$^2$ while the band
calculations predicted $\gamma_0\sim 10$
mJ/mol K$^2$ for all the three compounds.\cite{Nagao,bandmass} 
Thus, the experimental mass enhancement
factor $\gamma_{\rm K}/\gamma_0$ in KOs$_2$O$_6$ is 1.5-1.7 times larger than in
RbOs$_2$O$_6$ and CsOs$_2$O$_6$\cite{ele-ele}. 
In our calculation, $z_{\rm tot}^{\rm Rb}/z_{\rm
tot}^{\rm K}\simeq 1.71$ and $z_{\rm tot}^{\rm Cs}/z_{\rm
tot}^{\rm K}\simeq 1.98$, which are semi-quantitatively consistent with the 
experimental values. 
From our results, the enhanced mass enhancement in KOs$_2$O$_6$ is 
attributed to the proximity to the crossover to the quantum tunneling
state and also the 
large oscillation amplitude, which effectively enhances the
electron-phonon coupling. 

In order to obtain the correct value of the 
mass enhancement ($\sim 7$ for KOs$_2$O$_6$ and $\sim 4$ for the other
two), we should also take into account the electron-electron 
interactions, but this is beyond the scope of the present study and we
leave it for a future problem. 
Nevertheless, it is important to note that 
$\lambda$ is significantly enhanced near $b=b^*$ by anharmonic
oscillations and that this is the central
reason why the mass enhancement in KOs$_2$O$_6$ is much larger than in
 Rb and Cs compounds. 

As for the phonon frequency $\Delta$, 
 optical modes related to the $A$-cation
oscillations are observed at energy around 5-7 meV at room
temperature in the inelastic neutron
scattering\cite{NeutronSasai,Mukta} and their energies show strong 
softening as the temperature decreases. This
corresponds to our results shown in Fig. \ref{fig-Spectralfunc}.
For KOs$_2$O$_6$, the phonon energy at $T=1.5$ K is about 3.1 meV, 
which is quite small compared with the value of 5.5 meV at $T=300$ K.\cite{Mukta}
Furthermore, the low-energy phonon peaks 
in KOs$_2$O$_6$ are rather broad compared
with Rb and Cs compounds,\cite{Mukta} which
suggests the proximity to the crossover point.
The specific-heat experiment\cite{SummaryHiroi} and
 the photo-emission spectroscopy data\cite{PES} also support these
 results. 


In our calculation for KOs$_2$O$_6$ with parameters in Table \ref{tbl-1}, 
the peak position in the 
phonon spectrum is about 70 K $\simeq$ 6 meV at $T=300$ K and 
this shifts to $\Delta=38.5$ K as $T$ decreases. 
This means the  
potential parameters in Table \ref{tbl-1} can reproduce not only 
$T_{\rm c}$ and $\Delta$ but also the temperature dependence of 
the spectra.

As for the chemical trends, we have calculated superconducting
transition temperature for the three $\beta$-pyrochlore compounds 
adjusting parameters in the potential, Eq. (\ref{eq:pot}), and we can 
 reproduce the phonon energy and amplitude related to the average
Debye-Waller factor.
We have obtained quantitatively consistent values 
of $T_{\rm c}$ with the experimental ones, using 
 the same electron-phonon coupling, the electron DOS and also
the same additional high-energy phonon for all the three compounds.

\begin{table}[tb]
\caption{Local potential parameters and $\langle x^2\rangle$ at
 $T\!=\!0$, $\lambda$ at $T=T_{\rm c}$, $\Delta$, $T_{\rm c}$, and
 $T_{\rm c}^{(2)}$ for the three $\beta$-pyrochlore compounds. The other
 parameters are $B=9324\
 {\rm K/\AA}^3$, $C_2=C_1/4=3332$ K/\AA$^4$,  
and $\alpha^2=1897\ {\rm K/\AA}^2$. }
   \begin{tabular}{ccccccccc}
       & $\hbar\Omega_0({\rm K})$  & $b/b^*$ & $\langle x^2\rangle(T=0)$(\AA$^2$) & $\lambda$ 
     &  $\Delta$({\rm K})
     & $T_{\rm c}$({\rm K}) & $T_{\rm c}^{(2)}$(K)\\
         \hline
         \hline
K & 26.4 &   0.58 & 0.0152 & 1.47 & 38.5 & 6.45 & 10.5 \\ 
Rb & 54.6 &  0.28 & 0.0050 & 0.34 & 56.2 & 0.82& 5.74 \\ 
Cs & 74.8 &  0.11 & 0.0024 & 0.12 & 75.2 & $<$0.03& 3.37 \\ 
\hline
\hline
   \end{tabular}
\vspace{-0.3cm}
\label{tbl-1}
\end{table}

\subsection{Changes at $T_p$}\label{Tptran}
As mentioned in Sec. \ref{intro}, KOs$_2$O$_6$ exhibits an isomorphic 
first-order transition at $T_p=7.5$ K.\cite{SummaryHiroi,Hasegawa,newSasai} 
In Sec. \ref{Tptran}, let us investigate
the effects of this transition 
 on the phonon dynamics, using a set of constraints  
offered by the experimental results. 
We also discuss its effects 
on superconductivity.

Electric resistivity shows a concave temperature
dependence at high temperatures, 
and this is attributed to the strong 
coupling to phonons with strong 
anharmonicity.\cite{SummaryHiroi,DahmUeda} 
At the magnetic field $H=14$ T, the resistivity 
is suddenly suppressed by 25 \% at $T_p$, and 
shows a different $T$-dependence proportional to 
$T^2$ at $T_{\rm c}<T<T_p$.\cite{SummaryHiroi}  
This indicates that the electron-phonon 
scattering processes are reduced significantly at
$T_p$\cite{SummaryHiroi}.   
 This is also consistent with the reduction in the
specific-heat jump at $T_{\rm c}$ in magnetic
fields above 8 T.\cite{SummaryHiroi}

\begin{figure}[tb]
\begin{center}
    \includegraphics[width=0.44\textwidth]{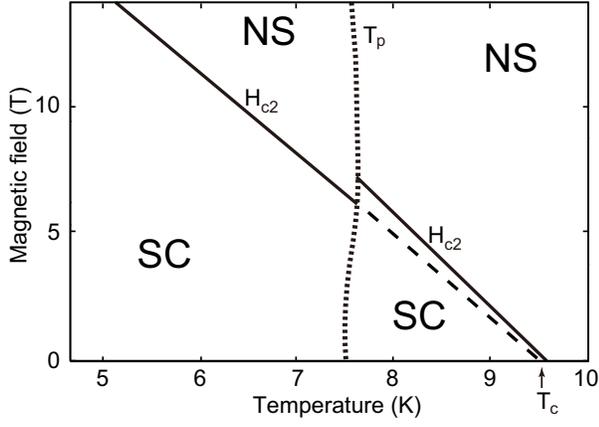}
\end{center}
\caption{Schematic temperature-magnetic field phase diagram taken from 
 Ref. 3. The solid and dotted lines represent $H_{c2}$ and $T_p$,
 respectively. The dashed line indicates the extrapolation of low-$T$
 $H_{c2}$ to the higher-temperature region.}
\label{fig-schematicHc2}
\end{figure}

In the magnetic field-temperature phase diagram, $T_p$ remains
 essentially insensitive to 
 magnetic field $H$ as shown in Fig. \ref{fig-schematicHc2}. 
The upper critical magnetic field $H_{c2}$ is suppressed 
 below $T_p$, but extrapolating to the
 region above $T_p$, it seems that it also vanishes at 
the position very close to $T_{\rm c}(H=0)$.

This fact cannot be explained if the characteristic 
energy scale, $\omega_{\rm log}$ or $\Delta$ is common across $T_p$: 
the extrapolated $H_{c2}$ line should vanish below $T_{\rm c}(H=0)$ 
since $\lambda$ is reduced below $T_p$ as discussed above. 
Therefore, the above fact implies the enhancement of the 
characteristic energy scale $\Delta$ or $\omega_{\rm log}$ below $T_p$. 
Indeed, Chang, {\it et al.},\cite{Chang} assumed a 
 slightly increased Einstein energy to fit the specific
heat data and our previous study also 
predicted the increase in the oscillation energy.\cite{Hat}

Let us discuss the changes in $\lambda$, $\Delta$, $|\langle
x^2\rangle|$, and $|\langle xyz \rangle|$ at $T_p$ based on our results, 
modeling the isomorphic transition by a
 sudden change in the two potential parameters, $\omega_0$ and $b$.
This change corresponds to the variation in the 
mean field part of inter-site ion interactions, which was 
discussed in Ref. 34 and also changes in the oxygen positions and the
lattice constant.\cite{newSasai}
Two parameters are chosen under the constraints 
(i) two different
parameter sets above and below $T_p$ 
lead to the same $T_{\rm c}$, and (ii) $\lambda$ is
smaller below $T_p$, both of which are the experimental constraints, and
we also assume (iii)
 no significant change in the electron
band structure across the transition and thus $\alpha^2$ is also
unchanged, and 
(iv) $b<b^*$ for both below and above $T_p$ implied by the
result of the electron-density profile obtained by the x-ray and the neutron
experiments 
\cite{Yamaura,newSasai} as discussed before.
In the following, we discuss the change across 
$T_p$ based on these constraints.

\begin{figure}[t]
\begin{center}
    \includegraphics[width=0.44\textwidth]{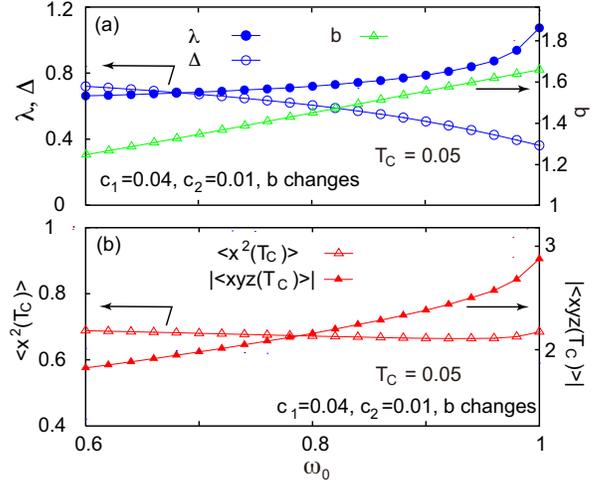}
\end{center}
\caption{(Color online) $\omega_0$ dependence of characteristic quantities, when $T_{\rm c}$ 
is fixed to 0.05 by tuning $b$.
(a) $\lambda$, $\Delta$, and $b$. (b) The second and 
third-order moments of ion oscillation $\langle
 x^2\rangle$, and $|\langle xyz\rangle|$.  $c_1=0.04$ and $c_2=0.01$.}
\label{fig-discussion}
\end{figure}
\begin{figure}[b]
\begin{center}
    \includegraphics[width=0.44\textwidth]{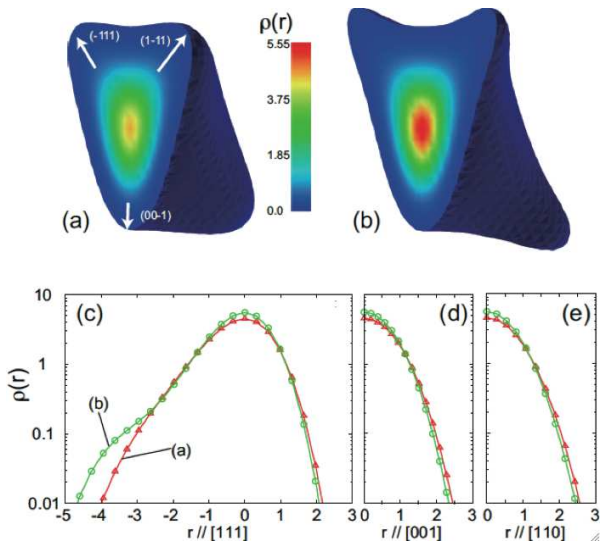}
\end{center}
\caption{(Color online) 
Ion density isosurface for $\rho({\bf r})=0.05$ and the density map on the
 section with (110) plane at $T=0.002$. (a) $\omega_0=0.8$, $b=1.44$ and
 (b) $\omega_0=1$, $b=1.66$. 
(c) $\rho({\bf r})$ along [111] direction corresponding to (a) and (b),
 (d) along [001] and (e) along [110] direction. $\rho(-|{\bf r}|)$'s in
 (d) and (e) are not shown since $\rho({\bf r})=\rho(-{\bf r})$
 along these directions.} 
\label{fig-3dfig}
\end{figure}

In Fig. \ref{fig-discussion}, characteristic quantities are shown for
fixed $c_1$ and $c_2$ with varying  $\omega_0$ and simultaneously $b$
such that those  
give the same $T_{\rm c}=0.05$. Since $b<b^*$ 
 which is the assumption (iv), $T_{\rm c}$ is well approximated by the McMillan formula
(\ref{McMillan}), in which $T_{\rm c}$ is determined by the two factors
$\omega_{\rm log}\simeq \Delta$ and $\lambda$.

Firstly, it is noted that
$\lambda$ increases as $\omega_0$ increases. One might expect a 
suppression of $\lambda$ as $\omega_0$ increases, but the
key point is that we simultaneously tune $b$ to fix $T_{\rm c}$ 
unchanged and  
this means larger $b$ is necessary at larger $\omega_0$. 
This increase in $b$ overcomes the competing effect of the 
increase in $\omega_0$, and finally $\lambda$ increases. 
Secondly, the energy of the first excited state $\Delta$ increases as 
$\omega_0$ decreases. This is 
naturally understood by noting that $\Delta$ becomes large for smaller
$b$. Thus, from the constraint  (ii) the above results imply 
 that the phonon energy is enhanced below $T_p$ 
and, indeed, this is consistent with the 
previous theoretical studies.\cite{Chang,Hat}
%
%

Most interestingly, the second- and third-order fluctuations 
 $\langle x^2\rangle$ and
$|\langle xyz\rangle|$ change oppositely as shown in
 Fig. \ref{fig-discussion}(b). As $\omega_0$ decreases, 
$\langle x^2\rangle$ increases for the most of
the $\omega_0$ range, while $|\langle xyz \rangle|$
decreases monotonically. This is  because 
small $\omega_0$ corresponds to 
small $b$ as explained above. 
These results indicate that
$\langle x^2\rangle$ is slightly enhanced below $T_p$ while 
$|\langle xyz \rangle|$ is suppressed.

In order to illustrate these anisotropic fluctuations,  
the density isosurface for $\rho({\bf r})=0.05$ and the 
density (color) map are shown 
for two different
parameters at $T=0.002$ in Fig. \ref{fig-3dfig}: 
%
%
for (a) and (b), $\omega_0=0.8$ and $1.0$, 
and $b=1.44$ and $1.66$, respectively, and these two sets give the same 
$T_{\rm c}=0.045$, and  we choose the two parameter sets 
to emphasize the change 
 in the oscillation profile. As expected from the values of $b$, the case (a) has the smaller 
$\lambda=0.647$ and the larger $\Delta=0.662$, while $\lambda=0.782$ and 
$\Delta=0.429$ for the case (b). 

%
%
%
%
One can see that the density
 isosurface shows more anisotropic character in the case (b), 
where the ``spikes'' sticking out 
along four [111] directions are
sharper than those in the case (a). This aspect reflects in 
the values of $\langle x^2 \rangle$ and $\langle xyz \rangle$.
Indeed, 
$\langle x^2 \rangle$ 
 at $T=T_{\rm c}$ is larger $\langle x^2 \rangle =0.645$ in (a) than $0.626$ in (b), while 
$|\langle xyz \rangle|$ at $T=T_{\rm c}$ is smaller $|\langle xyz\rangle|=1.93$ in (a) than $2.35$ in (b).
Figures \ref{fig-3dfig}(c)-\ref{fig-3dfig}(e) show
$\rho({\bf r})$ along [111], [001], and [110] directions, respectively, 
for the same two data. It is clearly
 seen that $|\langle xyz \rangle|$ in (b) is larger than that in (a)
as shown in (c), 
and $\langle x^2\rangle$ in (a) is larger than in (b) 
 as shown in (d) and (e). 

These changes in the anisotropy of 
density distribution is a 
 characteristic nature of the first-order isomorphic
 transition at $T_p$. This transition is isomorphic, consistent with the
 experiments,\cite{SummaryHiroi,Hasegawa} 
 because the amplitude of anisotropy changes but it does not break the 
point group, translational, and any other symmetries. 
 Our results suggest that $\langle x^2\rangle$
is slightly enhanced while $|\langle xyz \rangle|$ is suppressed below $T_p$ in KOs$_2$O$_6$. 
Recent high-resolution neutron-scattering experiment shows the increase
in $\langle x^2\rangle$ across $T_p$ as $T$ decreases,\cite{newSasai} 
 which is consistent with our result.

Furthermore, although it is naively expected that an
increase in $\langle x^2\rangle$ corresponds to the enhancement of 
$\lambda$, this does not
necessarily hold in this system, since the value of $\lambda$ is
sensitive to the value of $b$, especially near the crossover point $b^*$.
Our results show that the value of $|\langle xyz \rangle|$ plays more 
important role for the enhancement of $\lambda$ than $\langle x^2\rangle$.
In order to detect this change in the density distribution 
across $T_p$, it is important to perform detailed 
neutron-scattering experiments and to analyze the results with 
taking into account the third-order term in
Debye-Waller factor as discussed in Sec. \ref{debye-waller}.  
It is worthwhile to examine the ${\bf Q}$ dependence of the Debye-Waller
factor to extract the anisotropy of the oscillations, since 
${\bf Q}$-averaged Debye-Waller factor hinders the anisotropy and anharmonicity.

Finally, we comment on the possibility of re-entrant
superconducting transition. 
Experimentally, 
Hiroi {\it et al.},\cite{SummaryHiroi} 
observed a re-entrant superconducting transition 
around $H=7$ T as 
shown in Fig. \ref{fig-schematicHc2}. 
This can be understood from $H_{c2}(T=0)$ in the strong
coupling theory of superconductivity:
\begin{eqnarray}
H_{c2}(T=0)\propto T_{\rm c}^2(1+\lambda)^2(1+1.44\frac{T_{\rm c}}{\omega_{\rm log}}+\cdots),
\label{Hc2}
\end{eqnarray}
for the clean limit.\cite{Carbotte} 
In our analysis, the $\lambda$ decreases and
$\omega_{\rm log}$ increases below $T_p$, 
while the $T_{\rm c}$ is approximately the same
for the potential parameters above and below $T_p$.
Thus,  $H_{c2}(T=0)$ for the low-$T$ parameter set
 is smaller than that for the high-$T$ set. 
Therefore, we expect a $H_{c2}(T)$ curve similar to that shown in
Fig. \ref{fig-schematicHc2}. 
This result does not change when we
use the formula for the dirty limit.\cite{Carbotte}

\section{Summary}\label{conclusion}
We have investigated in the present paper anharmonic phonons and strong coupling
superconductivity in $\beta$-pyrochlore compounds, $A$Os$_2$O$_6$ ($A$=K,
Rb, or Cs).

First, we have solved the Schr\"odinger equation of the three-dimensional
anharmonic phonon in tetrahedral symmetry. The main issue is the
 importance of the third-order anharmonicity $b xyz$ in the ion potential 
allowed for this symmetry.  We have determined the
energy spectrum of the anharmonic phonon as a function of the 
third-order anharmonicity $b$. We have found that there exists a
crossover of the ground state to the quantum tunneling state for $b>b^*$.  
 We have pointed out non-monotonic temperature dependence of ion 
density profile near the crossover point $b^*$. 

Secondly, we have calculated the transition temperature of
superconductivity $T_{\rm c}$ mediated by these anharmonic phonons. 
We have found that the 
 enhancement of $T_{\rm c}$ near $b=b^*$. Its $b$ dependence can be
well fitted by the McMillan formula at $b\simle b^*$. For $b\simge 
b^*$ we have found that there exists an additional channel of
pairing interaction, which turns out to originate from low-energy 
excited states appearing at $b\sim b^*$. We
have analyzed its contribution and derived an
 approximated formula of $T_{\rm c}$ in the strong coupling limit. 

We have also discussed the chemical trends of $T_{\rm c}$ in the 
$\beta$-pyrochlore family 
$A$Os$_2$O$_6$ ($A$=K, Rb, or Cs). The main difference among the 
three members are different values of $b/b^*$ and 
the second-order term in the ion potential $\omega_0$.  
By assuming the same electron-phonon coupling constant $\alpha^2$
 and the same high-energy phonon for all the three compounds,
the differences in $T_{\rm c}$ and the energy of the 
first excited phonon states 
 $\Delta$ have been quantitatively explained 
only by the difference in the local anharmonic potential.

Finally, we have discussed the effect of the first-order transition observed in 
KOs$_2$O$_6$. The changes in the density distribution of K cation 
at the first-order transition have been discussed based on the
experimental data obtained so far. Especially, 
 we have found that $\langle x^2 \rangle$ and
 $|\langle xyz \rangle|$ change differently across $T_p$. Our results
 suggest the increase in the first excited phonon energy $\Delta$ and
 the reduction in the 
 dimensionless electron-phonon coupling  constant $\lambda$ 
 across $T_p$. While the latter is consistent with the experimental
results, the former has not been observed and this is our prediction 
for the experiments. In order to
detect the increase in $\Delta$, it is important to carry out the high-resolution inelastic
neutron scattering experiments. 
 We hope our results shed lights in $\beta$-pyrochlore compounds 
both on the strong coupling
superconductivity and on the anharmonic oscillations observed.

\begin{acknowledgements}
The authors would thank T. Dahm, K. Ueda, J. Yamaura and Z. Hiroi for
grateful discussions. They also acknowledge the international workshop ``{\it New
Developments in Theory of Superconductivity}'' held at Institute for
Solid State Physics, University of Tokyo, during June 22-July 10, 2009 for
providing an opportunity of discussion with other participants. 
This work is supported by KAKENHI (Grants No. 19052003 and No. 20740189) and
 also by the Next Generation Super Computing Project, Nanoscience 
Program, from the MEXT of Japan. 
\end{acknowledgements}


\begin{thebibliography}{99}
\bibitem{omega0var} In those cases of $\omega_0\ne 1$, we keep the same
	energy unit of 44 K, instead of renormalization of energy. 
This simplifies our discussions about $\omega_0$ dependence in later sections.
\bibitem{dimension} When $\lambda$ is expressed by the quantities with
	the dimension, $\lambda=(2\alpha^2/\Delta)(\hbar/2M\Delta)$, which
	is, indeed, dimensionless.
\bibitem{NoteBandCalc} The result in Ref. 33 suggests a slightly 
negative value of $\omega_0$ for KOs$_2$O$_6$. 
However, the first excited state then has an energy too small compared with 
the experimental estimate from the 
specific heat data (Ref. 3) and the neutron
	scattering (Refs. 25 and 27).  Correspondingly, calculated $\langle x^2\rangle$ is
	too large compared with the experimental results and its
	temperature dependence is qualitatively different from  the
	experimental one (Refs. 25 and 26). 
In this paper, we consider the cases of $\omega_0 \ge 0$ for
	simplicity. We hope that more detailed information on the ion
	potential 
        is provided by the neutron scattering experiments in the
	future.
\bibitem{InThisPoint} In this point whether the additional phonon is
	acoustic or optical mode does not make an essential difference.  
To improve estimation of $T_{\rm c}$, we 
 need the detailed information of each compound on the
electron-phonon coupling $\alpha^2$, $\Omega_h$, $\lambda_h$, and other parameters.
\bibitem{ele-ele} Note that this value of the mass enhancement factor includes not
	only the contributions of electron-phonon interactions but also 
	those of electron-electron interactions.
\end{thebibliography}
\end{document}